\documentclass[iop]{emulateapj}

\usepackage{amsmath}
\usepackage{multirow}
\usepackage{hhline}
\usepackage{hyperref}
\shorttitle{GRB 081024B and GRB 140402A}
\shortauthors{Aimuratov et al.}

\begin{document}

\title{GRB 081024B and GRB 140402A: two additional short GRBs from binary neutron star mergers}

\author{Y. Aimuratov\altaffilmark{1,2}, R.~Ruffini\altaffilmark{1,2,3,4}, M.~Muccino\altaffilmark{1,3}, C.~L.~Bianco\altaffilmark{1,3}, A.~V.~Penacchioni\altaffilmark{3,5,6}, G.~B.~Pisani\altaffilmark{1,3}, D.~Primorac\altaffilmark{1,3}, J.~A.~Rueda\altaffilmark{1,3,4}, Y.~Wang\altaffilmark{1,3}}

\altaffiltext{1}{ICRA and Dipartimento di Fisica, Sapienza Universit\`a  di Roma and ICRA, Piazzale Aldo Moro 5, 00185 Roma, Italy}
\altaffiltext{2}{Universit\'e de Nice Sophia-Antipolis, Grand Ch\^ateau Parc Valrose, Nice, CEDEX 2, France} 
\altaffiltext{3}{ICRANet, P.zza della Repubblica 10, 65122 Pescara, Italy}
\altaffiltext{4}{ICRANet-Rio, Centro Brasileiro de Pesquisas F\'isicas, Rua Dr. Xavier Sigaud 150, 22290--180 Rio de Janeiro, Brazil}
\altaffiltext{5}{ASI Science Data Center, Via del Politecnico s.n.c., 00133 Rome, Italy}
\altaffiltext{6}{Dept. of Physical Sciences, Earth and Environment, University of Siena, Via Roma 56, 53100 Siena, Italy}

\begin{abstract}
Theoretical and observational evidences have been recently gained for a two-fold classification of short bursts: 1) short gamma-ray flashes (S-GRFs), with isotropic energy $E_{iso}<10^{52}$~erg and no BH formation, and 2) the authentic short gamma-ray bursts (S-GRBs), with isotropic energy $E_{iso}>10^{52}$~erg evidencing a BH formation in the binary neutron star merging process. The signature for the BH formation consists in the on-set of the high energy ($0.1$--$100$~GeV) emission, coeval to the prompt emission, in all S-GRBs. No GeV emission is expected nor observed in the S-GRFs. In this paper we present two additional S-GRBs, GRB 081024B and GRB 140402A, following the already identified S-GRBs, i.e., GRB 090227B, GRB 090510 and GRB 140619B. We also return on the absence of the GeV emission of the S-GRB 090227B, at an angle of $71^{\rm{o}}$ from the \textit{Fermi}-LAT boresight. All the correctly identified S-GRBs correlate to the high energy emission, implying no significant presence of beaming in the GeV emission.
The existence of a common power-law behavior in the GeV luminosities, following the BH formation, when measured in the source rest-frame, points to a commonality in the mass and spin of the newly-formed BH in all S-GRBs.
\end{abstract}


\maketitle

%%%%%%%%%%%%%%%%%%%%%%%%%%%%
\section{Introduction}\label{sec:1}
%%%%%%%%%%%%%%%%%%%%%%%%%%%%

Gamma-ray bursts (GRBs) have been historically divided into a two-fold classification based on the observed $T_{90}$ duration of their prompt emission: short GRBs with $T_{90}\lesssim2$~s and long GRBs with $T_{90}\gtrsim2$~s \citep{1981Ap&SS..80....3M,Klebesadel1992,Dezalay1992,Koveliotou1993,Tavani1998}.

The progenitor systems of short bursts are traditionally identified with binary neutron star (NS)and NS-black hole (BH) mergers \citep[see, e.g.,][]
{Goodman1986,Paczynski1986,Eichler1989,Narayan1991,MeszarosRees1997_b,Rosswog2003,Lee2004,2006ApJ...648.1110B,2014ARA&A..52...43B}.
This assumption has received observational supports by their localization, made possible by the X-ray emission of the afterglow, with large off-sets from their hosts galaxies, both late and early type galaxies with no star formation evidence \citep[see, e.g.,][]{Fox2005,Gehrels2005,2014ARA&A..52...43B}.

A vast activity of numerical work on relativistic magnetohydrodynamical (MHD) simulation using the largest facilities in the world (equipped by supercomputers with peak performances of $6.8$ PFLOPS\footnote{The acronym PFLOPS means Peta ($10^{15}$) Floating Point Operations per second.}, see \citealt{2014ApJ...785L...6S}, $13.3$ PFLOPS, see \citealt{2016ApJ...824L...6R}, and $10.51$ PFLOPS, see \citealt{2014PhRvD..90d1502K}) have been developed with the declared goal of finding a jetted emission which they considered, without convincing observational support, to be a necessary step to develop short GRB models in merging binary NS-NS or binary BH-NS systems \citep[see, e.g.,][]{2011ApJ...732L...6R,2011ApJ...734L..36S,2014PhRvD..90d1502K,2014ApJ...785L...6S,2015ApJ...806L..14P,2016ApJ...824L...6R}. It is interesting that they themselves recognized the shortcoming of their approach: ``...there is microphysics that we do not model here, such as the effects of a realistic hot, nuclear EOS [equation of state] and neutrino transport'' \citep[see, e.g.,][]{2016ApJ...824L...6R}. They also expected such models would be further confirmed by the observation associated with gravitational waves (GWs) of aLIGO \citep[see, e.g.,][]{2004CQGra..21S1625B}.

There is no observational signature for the role of MHD activities in GRBs, nor, as we show in this paper, for jetted emission in the X- and $\gamma$-rays, as well as in the ultrarelativistic GeV emission of short bursts (see Sec.~\ref{sec:5}).
On the contrary, also in the case of short GRBs we have strong evidence for the necessary occurrence of hypercritical accretion process as already shown in long GRBs with the fundamental role of neutrino emission \citep{Zeldovich1972,RRWilson1973,2012ApJ...758L...7R,2014ApJ...793L..36F} and the value of the NS critical mass $M_{\rm crit}^{\rm NS}$ (\citealp{2011PhRvC..83d5805R,2011PhLB..701..667R,2011NuPhA.872..286R,2014PhRvC..89c5804R,2013IJMPD..2260007R,Belvedere2012,Belvedere2014,Belvedere2015,CCFRR2015}, see also \citealp{2014ApJ...793L..36F,2015ApJ...812..100B,2015PhRvL.115w1102F,2016ApJ...833..107B}). We also established firm upper limits on the observation of GWs from short GRBs by aLIGO \citep{Oliveira2014,2015ApJ...808..190R,2016arXiv160203545R}.

Our approach is markedly different from the traditional ones. Since \citet{Ruffini2001c,Ruffini2001,Ruffini2001a} we started:
\begin{itemize}
\item[a)] daily systematic and independent analyses of the GRB data in the X-, $\gamma$-rays and GeV emission from \textit{Beppo-SAX} \citep[see, e.g.,][]{2015mgm..conf...33F}, \textit{Swift} \citep{Barthelmy2005}, \textit{Fermi} \citep{Meegan2009}, \textit{Konus-WIND} \citep{1995SSRv...71..265A}, and \textit{AGILE} \citep{2009A&A...502..995T}. We extended our data analysis to the optical and radio data.
\item[b)] We have developed theoretical and astrophysical models based on quantum and classical relativistic field theories.
\item[c)] At every step we have verified that the theoretical considerations be consistent with the observational data.
\end{itemize}

In this article we mainly address the study of NS--NS mergers and only at the end we refer to BH--NS binaries.

In \citet{2015ApJ...808..190R} a further division of the short bursts into two different sub-classes has been proposed, and specific observable criteria characterizing this division have been there given:
\begin{enumerate}
\item The first sub-class of short bursts is characterized by isotropic energies $E_{\rm iso}\lesssim 10^{52}$~erg and rest-frame spectral peak energies $E_{p,i}\lesssim2$~MeV \citep{2012ApJ...750...88Z,Calderone2014}.
In this case the outcome of the NS--NS merger is a massive NS (MNS) with additional orbiting material \citep{2016ApJ...832..136R}. An alternative  scenario leads to a new binary system composed by a MNS and a less massive NS or a white dwarf (WD). For specific mass-ratios a stable mass-transfer process may occur from the less massive to the MNS \citep[see, e.g.,][and references therein]{1977ApJ...215..311C,1992ApJ...400..175B}. Consequently, the donor NS moves outward by loosing mass and may also reach the beta-decay instability becoming a low-mass WD. In view of their moderate hardness and their low energetics, we have indicated such short bursts as short gamma-ray flashes 
\citep[S-GRFs, see][]{2016ApJ...832..136R}. There, the local rate of S-GRFs has been estimated to be $\rho_0=3.6^{+1.4}_{-1.0}$~Gpc$^{-3}$~yr$^{-1}$.
\item The second sub-class corresponds to the authentic short GRBs (S-GRBs) with $E_{\rm iso}\gtrsim 10^{52}$~erg and $E_{p,i}\gtrsim2$~MeV \citep{2012ApJ...750...88Z,Calderone2014}.
In this system the NS--NS merger leads to the formation of a Kerr BH with additional orbiting material, in order to conserve energy and angular momentum \citep{2016ApJ...831..178R,2016ApJ...832..136R}.
A further characterizing feature of S-GRBs absent in S-GRFs is the presence of the $0.1$--$100$~GeV emission, coeval to their prompt emission and evidencing the activity of the newly-born BH. In \citet{2016ApJ...832..136R} the local rate of this S-GRB has been estimated to be $\rho_0=\left(1.9^{+2.8}_{-1.1}\right)\times10^{-3}$~Gpc$^{-3}$~yr$^{-1}$. The impossibility of detecting the observed short GRB 140619B from LIGO was evidenced \citep[see Fig.$~12$ in][]{2015ApJ...808..190R}. We return again in this article on the issue of non-detectability of GWs for S-GRBs.
\end{enumerate}

The above relative rate of these two sub-classes of short bursts has been discussed and presented in \citet{2016ApJ...832..136R}. There, it has been shown that the S-GRFs are the most frequent events among the short bursts. This conclusion is in good agreement with the NS--NS binaries 
observed within our Galaxy: only a subset of them has a total mass larger than $M_{\rm crit}^{\rm NS}$ and can form a BH in their merging process \citep[][]{2015ApJ...808..190R}.
There, in Fig.~$3$, it has been assumed $M_{\rm crit}^{\rm NS}=2.67M_\odot$ for a non-rotating NS, imposing global charge neutrality and using the NL3 nuclear model \citep[see, e.g.,][]{CCFRR2015}.
Similar conclusions have been also independently reached by \citet{2015arXiv150407605F} and \citet{2015ApJ...808..186L}.

We have identified three authentic S-GRBs: GRB 090227B \citep{Muccino2012}, GRB 090510 \citep{2016ApJ...831..178R}, and GRB 140619B \citep{2015ApJ...808..190R}. All of them populate the high energy part of the $E_{p,i}$--$E_{iso}$ relation for short bursts \citep{2012ApJ...750...88Z,Calderone2014,2016ApJ...831..178R}
and have $E_{iso}>10^{52}$~erg. We have analyzed the above three S-GRBs within the fireshell model \citep[see e.g.,][]{2010PhR...487....1R}. The transparency emission of the $e^+e^-$ plasma (the P-GRB emission), the on-set of the prompt emission, the correlation between the spike emission of the prompt and CBM inhomogeneities have led to the most successful test and applicability of the fireshell model.

A further and independent distinguishing feature between S-GRFs and S-GRBs has been found thank to the \textit{Fermi} data: when these three S-GRBs fall within the \textit{Fermi}-LAT field of view (FoV), a GeV emission occurs, starting soon after the P-GRB emission, related to the emission from a newly-born BH.

In this paper, we present two additional S-GRBs: GRB 081024B and GRB 140402A. 
The S-GRB 081024 is historically important since that source gave the first clear detection of a GeV temporal extended emission from a short burst  \citep{2010ApJ...712..558A}.
From the application of the fireshell model to this S-GRB we theoretically derived its redshift $z=3.12\pm1.82$ and, therefore, $E_{iso}=(2.6\pm1.0)\times10^{52}$~erg, $E_{p,i}=(9.6\pm4.9)$~MeV, and $E_{\rm LAT}=(2.79\pm0.98)\times10^{52}$~erg.
For the S-GRB 140402A, we theoretically derived a redshift $z=5.52\pm0.93$ which provides $E_{iso}=(4.7\pm1.1)\times10^{52}$~erg and $E_{p,i}=(6.1\pm1.6)$~MeV. A long-lived GeV emission within $800$~s has been reported \citep{2014GCN..16069...1B}. The total energy of the brightest GeV emission is $E_{\rm LAT}=(4.5\pm2.2)\times10^{52}$~erg.

We also updated the analysis of the GeV emission of the S-GRB 090227B. 
The apparent absence of the GeV emission has been already discussed in \citet{2015ApJ...808..190R}, recalling that this source was outside the nominal LAT FoV, and only photons in the LAT low energy (LLE) channel and a single transient-class event with energy above $100$~MeV were associated with this GRB \citep{Ackermann2013}.
A further updated analysis would indicate that, in view of the missing observations, in no way the absence of the GeV emission before $\sim40$~s in the source rest-frame can be inferred. 

From the analyses of the two additional S-GRB 081024B and S-GRB 140402A and the further check for the GeV emission associated to the S-GRB 090227B, we conclude that all S-GRBs correlate to the high energy emission implying no significance presence of beaming in the GeV emission.  

In Sec.~\ref{sec:3} we briefly recall the fireshell model and its implications for S-GRBs.
In Secs.~\ref{sec:2} and \ref{sec:4} we report the data analyses of the S-GRBs 081024B and 140402A, respectively, and show their theoretical interpretation within the fireshell model: from the theoretical inference of their cosmological redshift, their transparency emission parameters, to the details of the circumburst media where they occurred.
In Sec.~\ref{sec:5} we summarize the properties of the GeV emission of all S-GRBs and show the characteristic common power-law behavior of ther rest-frame $0.1$--$100$~GeV luminosity light curves. We discuss also the minimum Lorentz factor of the GeV emission $\Gamma^{\rm min}_{\rm GeV}$ obtained by requiring that the outflow must be optically thin to GeV photons (namely to the pair creation process), as well as its possible energy source, i.e., the matter accretion onto the new formed BH.
In Sec.~\ref{sec:6} we indicate that there is no evidence in favor or against a common behavior of the X-ray afterglows of the S-GRBs in view of the limited observations.
In Sec.~\ref{sec:usgrb} 
we shortly address the issue of the possible emission of short bursts from BH-NS binaries leading to the ultrashort GRBs \citep[U-GRBs, see][]{2015PhRvL.115w1102F,2016ApJ...832..136R}
In Sec.~\ref{sec:conclusions} we infer our conclusions.

%%%%%%%%%%%%%%%%%%%%%%%%%%%%%%%%%%%%%%%%%%%%%%%%%%%%%%
\section{The fireshell model}\label{sec:3}
%%%%%%%%%%%%%%%%%%%%%%%%%%%%%%%%%%%%%%%%%%%%%%%%%%%%%%

In the fireshell model \citep{Ruffini2001c,Ruffini2001,Ruffini2001a}, the GRB acceleration process consists in the dynamics of an optically thick $e^+e^-$ plasma of total energy $E_{e^+e^-}^{\mathrm{tot}}$ -- the \textit{fireshell}. 
Its expansion and self-acceleration is due to the gradual $e^+e^-$ annihilation, which has been described in \citet{RSWX2}. 
The effect of baryonic contamination on the dynamics of the fireshell has been then considered in \citet{RSWX}, where it has been shown that even after the engulfment of a baryonic mass $M_B$, quantified by the baryon load $B=M_Bc^2/E_{e^+e^-}^{\rm tot}$, the fireshell remains still optically thick and continues its self-acceleration up to ultrarelativistic velocities \citep{2007PhRvL..99l5003A,2009PhRvD..79d3008A}. 
The dynamics of the fireshell in the optically thick phase up to the transparency condition is fully described by $E^{tot}_{e^+e^-}$ and $B$ \citep{RSWX}.
In the case of long bursts, it is characterized by $10^{-4}\lesssim B<10^{-2}$ \citep{Izzo2012,Patricelli,Penacchioni2011,Penacchioni2013}, while for short bursts we have $10^{-5}\lesssim B\lesssim 10^{-4}$ \citep{Muccino2012,2015ApJ...808..190R,2016ApJ...831..178R}.

The fireshell continues its self-acceleration until the transparency condition is reached; then a first flash of thermal radiation, the P-GRB, is emitted \citep{RSWX2,RSWX,Ruffini2001}.
The spectrum of the P-GRB is determined by the geometry of the fireshell which is dictated, in turn, by the geometry of the pair-creation region. 
In the case of the spherically symmetric dyadosphere, the P-GRB spectrum is generally described by a single thermal component in good agreement with the spectral data \citep[see, e.g.,][]{Muccino2012,2015ApJ...808..190R}. 
In the case of an axially symmetric dyadotorus, the resulting P-GRB spectrum is a convolution of thermal spectra of different temperatures which resembles more a power-law spectral energy distribution with an exponential cutoff \citep{2016ApJ...831..178R}.

After transparency, the accelerated baryons (and leptons) propagates through the circum-burst medium (CBM).
The collisions with the CBM, assumed to occur in fully radiative regime, give rise to the prompt emission \citep{Ruffini2001}.
The spectrum of these collisions, in the comoving frame of the shell, is modeled with a modified BB spectrum, obtained by introducing an additional power-law at low energy with a phenomenological index $\bar{\alpha}$ which describes the departure from the purely thermal case \citep[see][for details]{Patricelli}.
The structures observed in the prompt emission of a GRB depend on the CBM density $n_{CBM}$ and its inhomogeneities \citep{Ruffini2004}, described by the fireshell filling factor $\mathcal{R}$. This parameter is defined as the ratio between the effective fireshell emitting area $A_{eff}$ and the total visible area $A_{vis}$ \citep{Ruffini2002,Ruffini2005}.
The $n_{CBM}$ profile determines the temporal behavior (the \textit{spikes}) of the light curve.
The observed prompt emission spectrum results from the convolution of a large number of modified BB spectra over the surfaces of constant arrival time for photons at the detector \citep[EQuiTemporal Surfaces, EQTS,][]{Bianco2005b,Bianco2005a} over the entire observation time. Each modified BB spectrum is deduced from the interaction with the CBM and it is characterized by decreasing temperatures and Lorentz and Doppler factors.

The duration and, consequently, the moment at which the burst emission stops are determined by the dynamics of the $e^+e^-$ plasma. The short duration is essentially due to the low baryon load of the plasma and the high Lorentz factor $\Gamma\approx10^4$ (see Fig.~2 in \citealt{Ruffini2001} and Fig.~4 in \citealt{Muccino2012}.

The description of both the P-GRB and the prompt emission, requires the appropriate relative spacetime transformation paradigm introduced in \citet{Ruffini2001a}: it relates the observed GRB signal to its past light cone, defining the events on the worldline of the source that is essential for the interpretation of the data. This requires the knowledge of the correct equations relating the comoving time, the laboratory time, the arrival time, and the arrival time at the detector corrected by the cosmological effects.

It is interesting to compare and contrast the masses, densities, thickness and distances from the BH of the CBM clouds, both in short and long bursts. In S-GRBs we infer CBM clouds with masses of $10^{22}$--$10^{24}$~g and size
of $\approx10^{15}$--$10^{16}$~cm, at typical distances from the BH of $\approx10^{16}$--$10^{17}$~cm (see Secs.~\ref{sec:2.2.2} and \ref{sec:4.2.2} and \citealt{2016ApJ...831..178R}), indeed very similar to the values inferred in long GRBs \citep[see, e.g.,][]{Izzo2012}. The different durations of the spikes in the prompt emission of S-GRBs and long bursts depend, indeed, only on the different values of $\Gamma$ of the accelerated baryons and not on the structure of the CBM: in long bursts we have $\Gamma\approx10^2$--$10^3$ \citep[see, e.g.,][]{Izzo2012}, while in S-GRBs it reaches the value of $\Gamma\approx10^4$ \citep[see, e.g.,][]{2016ApJ...831..178R} (see Secs.~\ref{sec:2.2.2} and \ref{sec:4.2.2}).

The evolution of an optically thick baryon-loaded pair plasma, is generally described in terms of $E_{e^+e^-}^{\rm tot}$ and $B$ and it is independent of the way the pair plasma is created. This general formalism can also be applied to any optically thick $e^+e^-$ plasma, like the one created via $\nu \bar{\nu}\leftrightarrow e^+e^-$ mechanism in a NS merger as described in \citet{Narayan1992}, \citet{SalmonsonWilson2002}, and \citet{Rosswog2003}.

Only in the case in which a BH is formed, an additional component to the fireshell emission occurs both in S-GRBs and in the binary-driven hypernovae
(BdHNe, long GRBs with $E_{\rm iso}>10^{52}$~erg, details in \citealt{2017arXiv170403821R} at the end of the P-GRB phase: the GeV emission observed by \textit{Fermi}-LAT and AGILE.
As outlined in this article, this component has a Lorentz factor $\Gamma>300$ and, as we will show in Sec.~\ref{sec:5}, it appears to have an universal behavior common to S-GRBs and BdHNe.
It is however important to recall that the different geometry present in S-GRBs and BdHNe leads, in the case of BdHNe, to the absorption of the GeV emission in some specific cases \citep{2017arXiv170403821R}.

%%%%%%%%%%%%%%%%%%%%%%%%%%%%%%%%%%%%%%%%%%%%%%%%%%%%%%
\section{The S-GRB 081024B}\label{sec:2}
%%%%%%%%%%%%%%%%%%%%%%%%%%%%%%%%%%%%%%%%%%%%%%%%%%%%%%

\subsection{Observations and data analysis}\label{sec:2.1}

The short hard GRB 081024B was detected on 2008 October 24 at 21:22:41 (UT) by the \textit{Fermi}-GBM \citep{2008GCN..8408....1C}. 
It has a duration $T_{90}\approx0.8$~s long and exhibits two main peaks, the first one lasting $\approx0.2$~s.
Its location $(RA,\,Dec)=(322^{\rm{o}}.9,\,21^{\rm{o}}.204)$ (J2000) is consistent with that reported by the \textit{Fermi}-LAT \citep{2008GCN..8407....1O}.
The LAT recorded $11$ events with energy above $100$~MeV within $15^{\rm{o}}$ from the position of the burst and within $3$~s from the trigger time \citep{2010ApJ...712..558A}. Emission up to $3$~GeV was seen within $\sim5$~s after the trigger \citep{2008GCN..8407....1O}.

GRB 081024B also triggered the \textit{Suzaku}-WAM, showing a double peaked light curve with a duration of $\sim0.4$~s \citep{2008GCN..8444....1H}.
\textit{Swift}-XRT began observing the field of the \textit{Fermi}-LAT $\sim70.3$~ks after the trigger, in Photon Counting (PC) mode for~$9.9$ ks \citep{2008GCN..8410....1G}. 
Three uncatalogued sources were detected within the \textit{Fermi}-LAT error circle \citep{2008GCN..8410....1G}, but a series of follow-up observations established that none of them could be the X-ray counterpart because they were not fading \citep{2008GCN..8416....1G,2008GCN..8454....1G,2008GCN..8513....1G}.

The above possible associations have been also discarded by the optical observations performed in the $R_c$-band \citep{2008GCN..8456....1F}.
Consequently, no host galaxy has been associated to this burst and, therefore, there no spectroscopic redshift has been determined.

\subsubsection{Time-integrated spectral analysis of the \textit{Fermi}-GBM data}\label{sec:2.1.1}

We analyzed the data from the \textit{Fermi}-GBM detectors, i.e., the NaI-n6 and n9 ($8$--$900$~keV) and the BGO-b1 ($0.25$--$40$~MeV), and LAT data \footnote{\url{http://fermi.gsfc.nasa.gov/ssc/data/analysis/documentation/Cicerone/}} in the energy range $0.1$ -- $100$~GeV. 
In order to obtain detailed \textit{Fermi}-GBM light curves we analyzed the \texttt{TTE} (Time-Tagged Events) files \footnote{\url{ftp://legacy.gsfc.nasa.gov/fermi/data/gbm/bursts}} with the \texttt{RMFIT}  package.~\footnote{\url{http://fermi.gsfc.nasa.gov/ssc/data/analysis/rmfit/vc\_rmfit\_tutorial.pdf}}

\begin{figure}
\centering
\includegraphics[width=\hsize,clip]{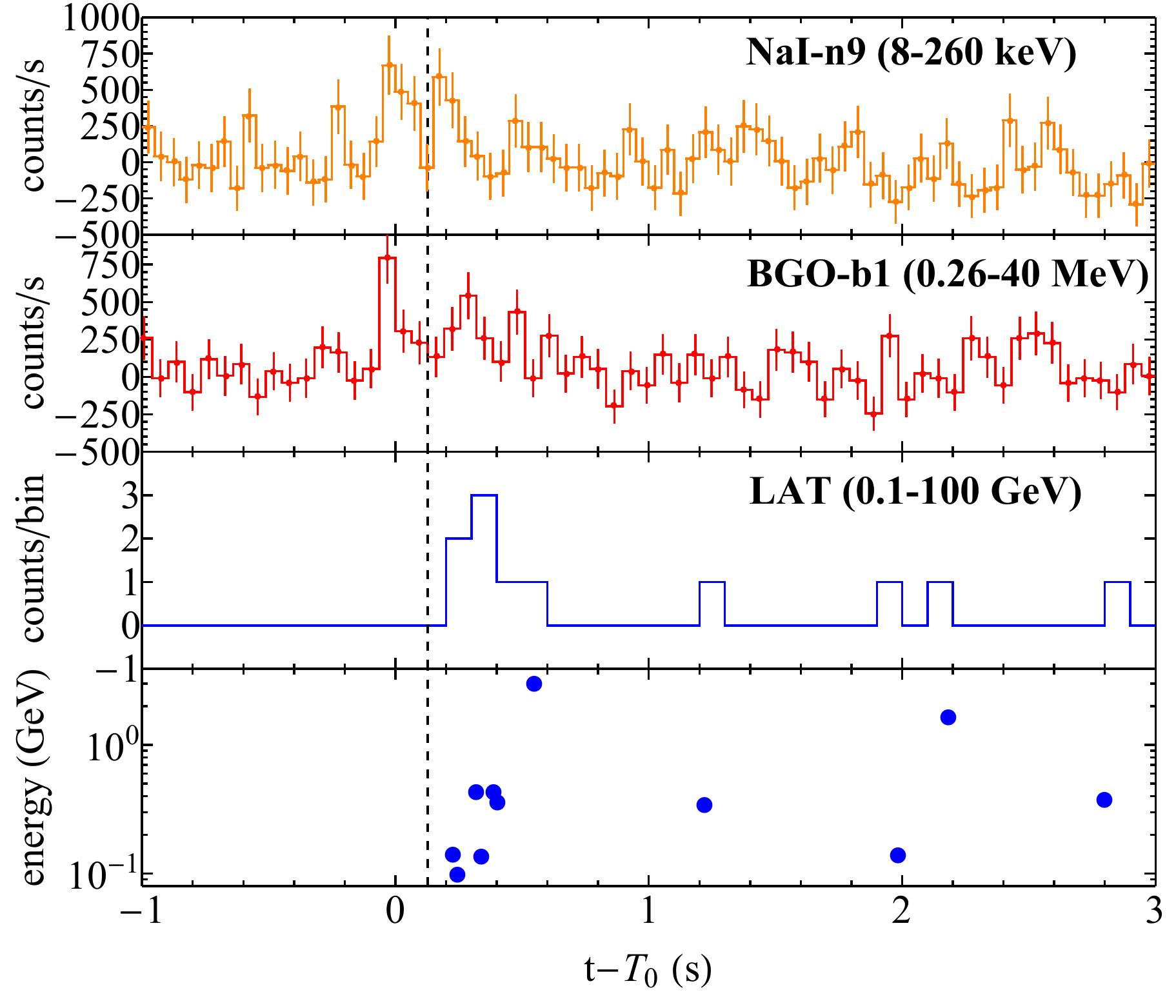}
\caption{Background subtracted light curves and high energy photons of GRB 081024B: the $50$~ms binned light curves from the NaI-n9 ($8$ -- $260$~keV, top panel) and BGO-b1 ($0.26$ -- $40$~MeV, second panel) detectors, the $100$~ms binned high-energy channel light curve ($0.1$ -- $100$ GeV, third panel, without error bars), and the high energy photons detected by the of the \textit{Fermi}-LAT (bottom panel). The vertical dashed line marks the end of the first \textit{Fermi}-GBM light curve pulse, before the on-set of the LAT light curve.}
\label{fig:f1a}
\end{figure}

In Fig.~\ref{fig:f1a} we reproduced the $50$~ms binned GBM light curves corresponding to the NaI-n9 ($8$ -- $260$~keV, top panel) and the BGO-b1 ($0.26$ -- $40$~MeV, second panel) detectors. We also reproduced the $100$~ms binned LAT light curve ($0.1$ -- $100$~GeV, third panel) and the corresponding high energy detected photons (bottom panel), both consistent with those reported in \citet{2010ApJ...712..558A}. All the light curves are background subtracted.
The GBM light curves show one narrow spike of about $0.1$~s, followed by a longer pulse lasting around $\sim0.7$~s.

\begin{figure*}
\centering
\includegraphics[height=0.49\hsize,clip,angle=90]{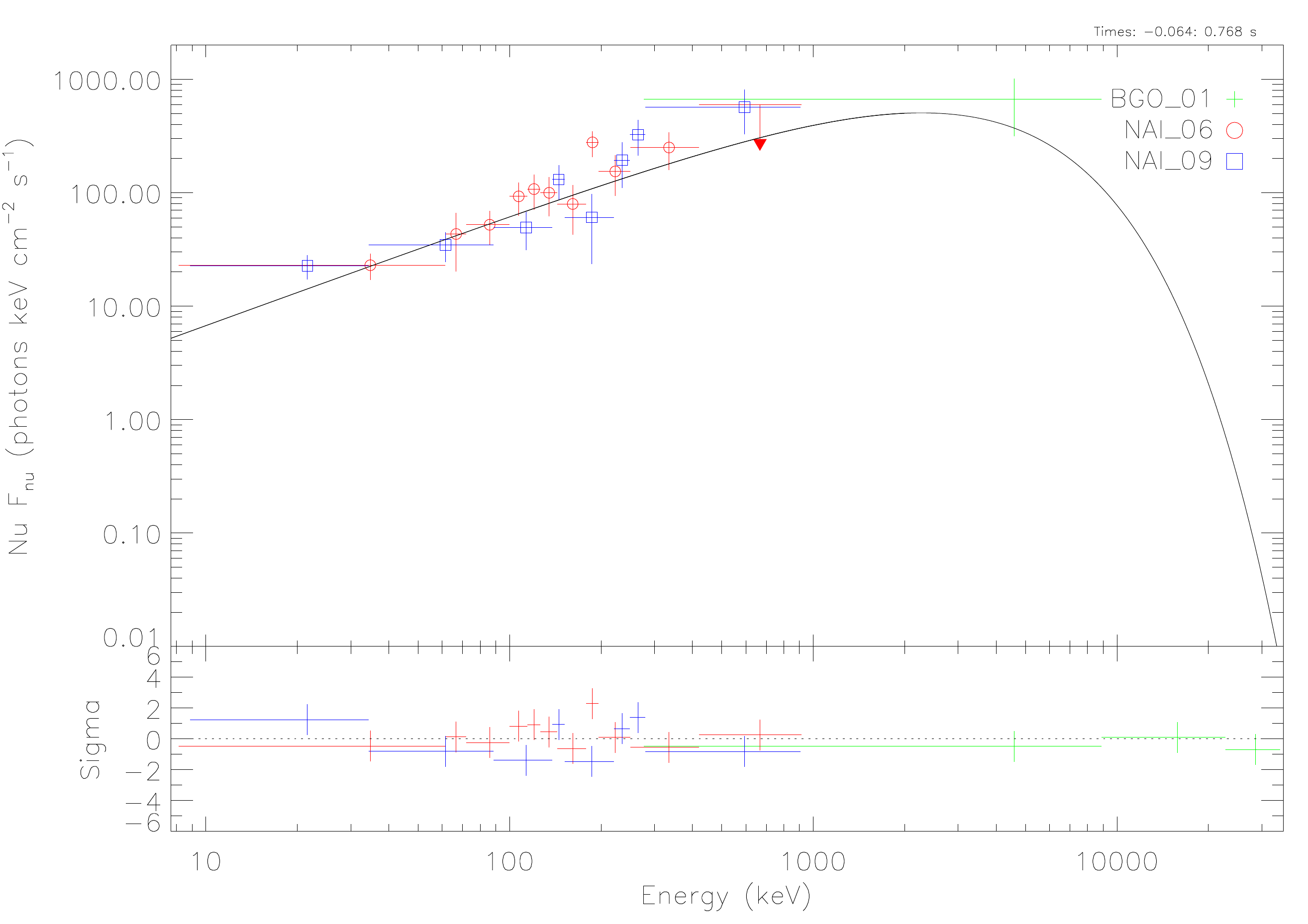}
\includegraphics[height=0.49\hsize,clip,angle=90]{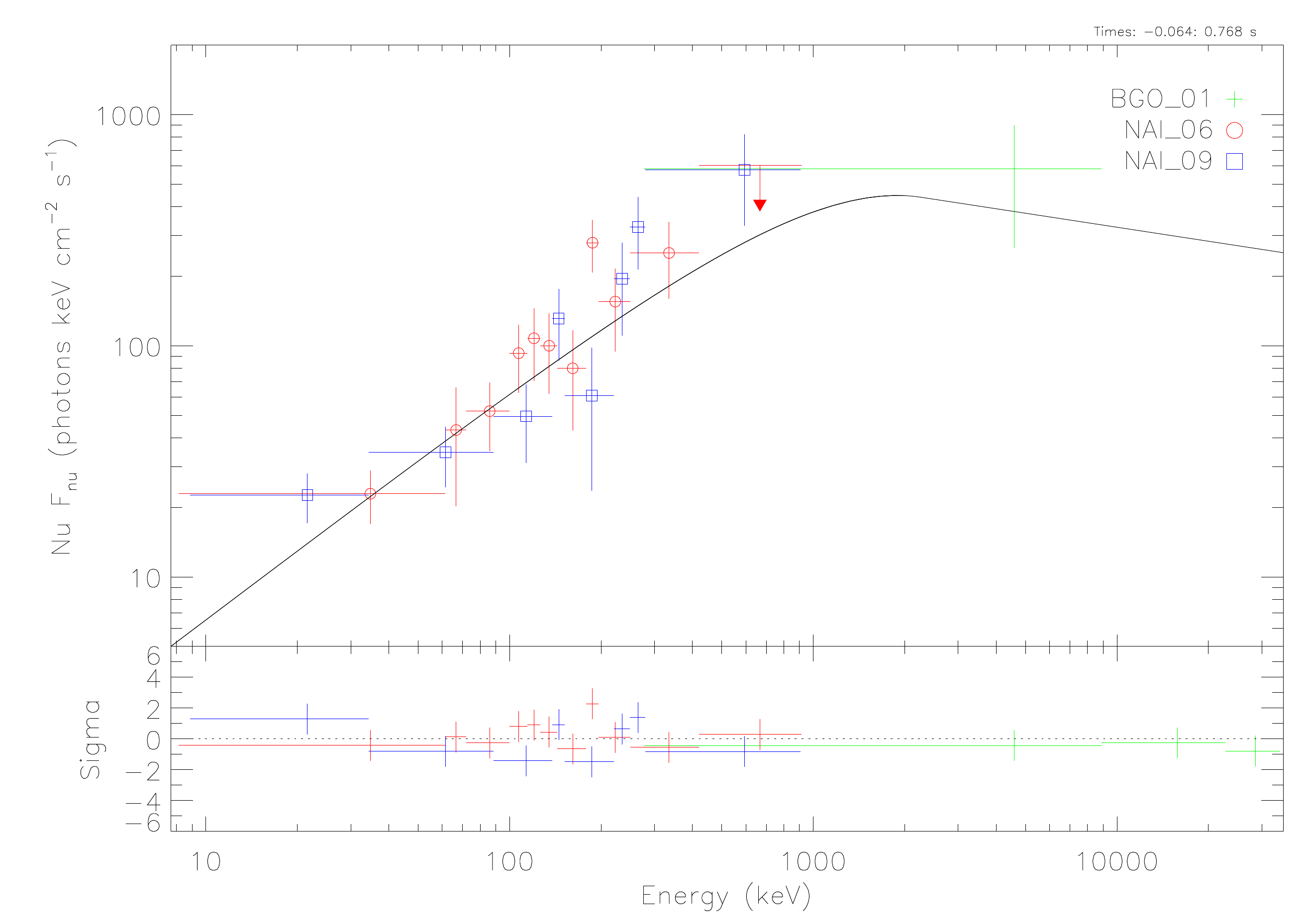}
\caption{The Compt (left plot) and the Band (right plot) spectral fits on the combined NaI--n6, n9+BGO--b0 $\nu F_\nu$ data of GRB 081024B in the $T_{90}$ time interval.}
\label{aba:fig1}
\end{figure*}

The time-integrated analysis was performed in the time interval from $T_0-0.064$ s to $T_0+0.768$ s which corresponds to $T_{90}$ duration of the burst and $T_0$ is the trigger time. 
We have fitted the corresponding spectrum with two spectral models: Comptonized (Compt, i.e., a power-law model with an exponential cutoff) and Band \citep{Band1993}, see Fig.~\ref{aba:fig1} and Tab.~\ref{tab:1}.
The Compt and the Band models provide similar values of the C-STAT (see Tab.~\ref{tab:1}). 
Therefore, the best-fit is the Compt model because it has one parameter less than the Band one.

\begin{table*}
\scriptsize
\centering
\begin{tabular}{cccccccccc}
\hline\hline
\multicolumn{9}{c}{S-GRB 081024B}\\
\hline
$\Delta T$          &  Model   &  $K$ (ph keV$^{-1}$ cm$^{-2}$s$^{-1}$)   &  $kT$ (keV)      &  $E_{p}$ (MeV)   &  $\alpha$  &  $\beta$ 
                        &  $F$ (erg cm$^{-2}$s$^{-1}$)   &  C-STAT/DOF    \\
\hline
$T_{90}$      &  Compt       &  $(6.39\pm0.69)\times10^{-3}$          &     &   $2.3\pm1.2$           & $-1.02\pm0.11$   &         &  $(2.27\pm0.87)\times 10^{-6}$                                                
                        &  $383.89/356$  \\
&                      Band   &  $(6.51\pm0.92)\times10^{-3}$                  &                       &  $1.9\pm1.7$        &  $-1.01\pm0.15$     &   $-2.2\pm1.1$     &  $(2.9\pm1.5)\times 10^{-6}$ 
                        &  $383.23/355$  \\
\hline
$\Delta T_{1}$    &  BB       &  $(4.0\pm1.7)\times10^{-7}$                  &  $152\pm20$    &                          &                    &       &  $(2.24\pm0.40)\times 10^{-6}$ 
                        &  $343.54/357$  \\
&                      Compt  &  $(9.8\pm1.9)\times10^{-3}$                      &                      &  $1.33\pm0.59$   &  $-0.48\pm0.27$    &     &  $(3.8\pm1.4)\times 10^{-6}$
                        &  $333.78/356$  \\
\hline
$\Delta T_{2}$    &  PL       &  $(4.60\pm0.53)\times10^{-3}$                     &      &                          &    $-1.37\pm0.07$          &            &  $(5.0\pm1.5)\times10^{-6}$
                        &  $392.2/357$  \\
&                      Compt  &  $(4.80\pm0.59)\times10^{-3}$                   &                       &  $10.95$(unc)    &   $-1.28\pm0.11$  &   &  $(3.5\pm2.0)\times10^{-6}$           
                        &  $390.57/356$  \\
\hline
\end{tabular}
\caption{Results of the spectral analyses on the S-GRB 081024B. Each column lists: the GRB, the time interval $\Delta T$, the spectral model, the normalization constant $K$, the BB temperature $kT$, the Compt peak energy $E_p$, the low-energy $\alpha$ and the high-energy $\beta$ photon indexes, the $8$ keV -- $40$ MeV energy flux $F$, and the value of the C-STAT over the number of degrees of freedom (DOF).}
\label{tab:1}
\end{table*}

%%%%%%%%%%%%%%%%%%%%%%%%%%%%%%%%%%%%%%%%%%%%%%%%%%%%%%
\subsubsection{Time-resolved spectral analysis of the \textit{Fermi}-GBM data}\label{sec:2.1.2}
%%%%%%%%%%%%%%%%%%%%%%%%%%%%%%%%%%%%%%%%%%%%%%%%%%%%%%

We have also performed the time-resolved analysis by using $16$ ms bins.
After the rebinning the GBM light curves still exhibit two pulses: the first pulse observed before the LAT emission on-set, from $T_0-0.064$~s to $T_0+0.128$~s, and the following emission, from $T_0+0.128$~s to $T_0+0.768$~s, hereafter dubbed as $\Delta T_1$ and $\Delta T_2$ time intervals, respectively.

As proposed in \citet{2015ApJ...808..190R}, the emission before the on-set of the LAT emission corresponds to the P-GRB emission, while the following emission is attributed to the prompt emission (see Sec.~\ref{sec:3}).

\begin{figure*}
\centering 
\includegraphics[height=0.49\hsize,clip,angle=90]{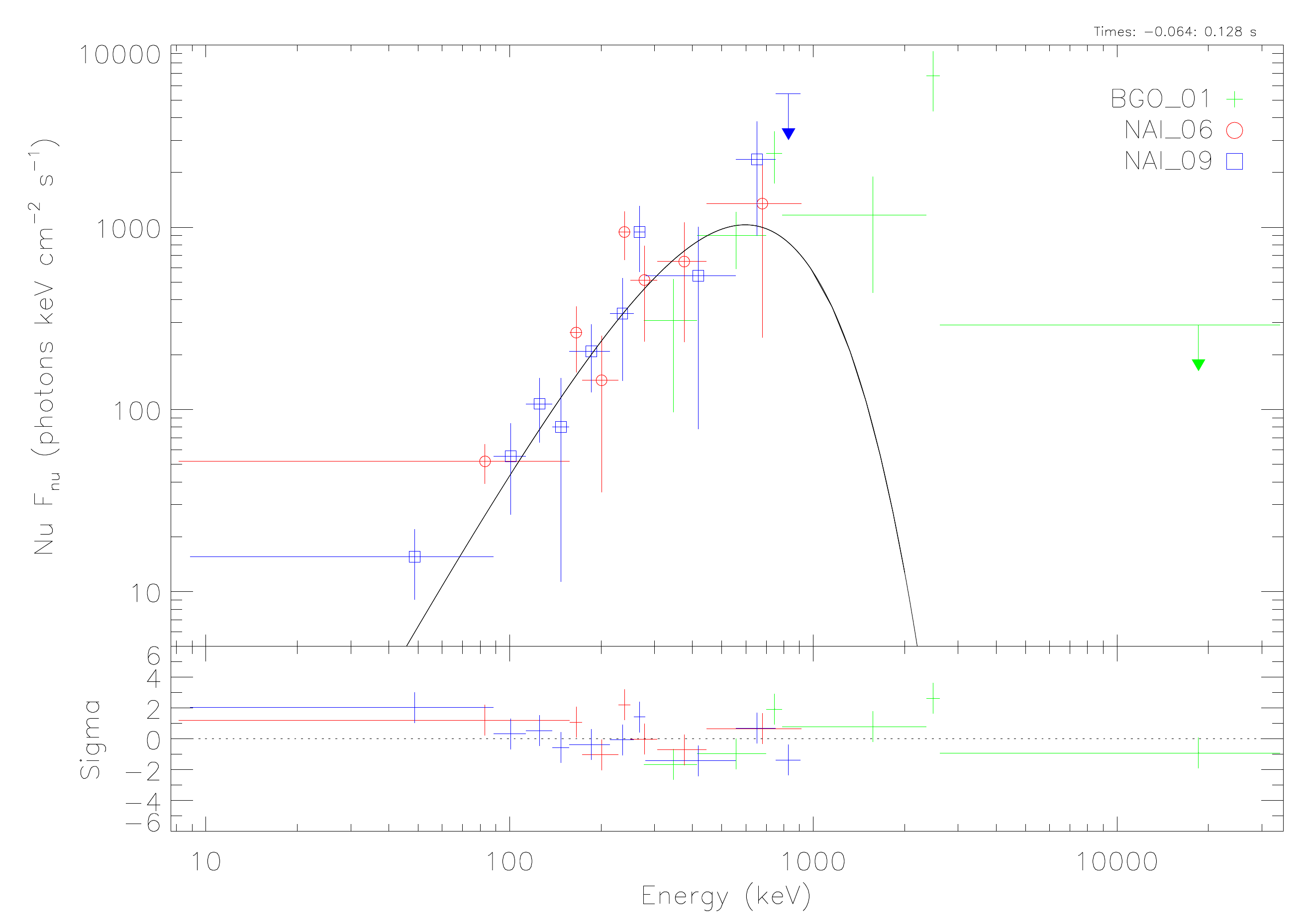}
\includegraphics[height=0.49\hsize,clip,angle=90]{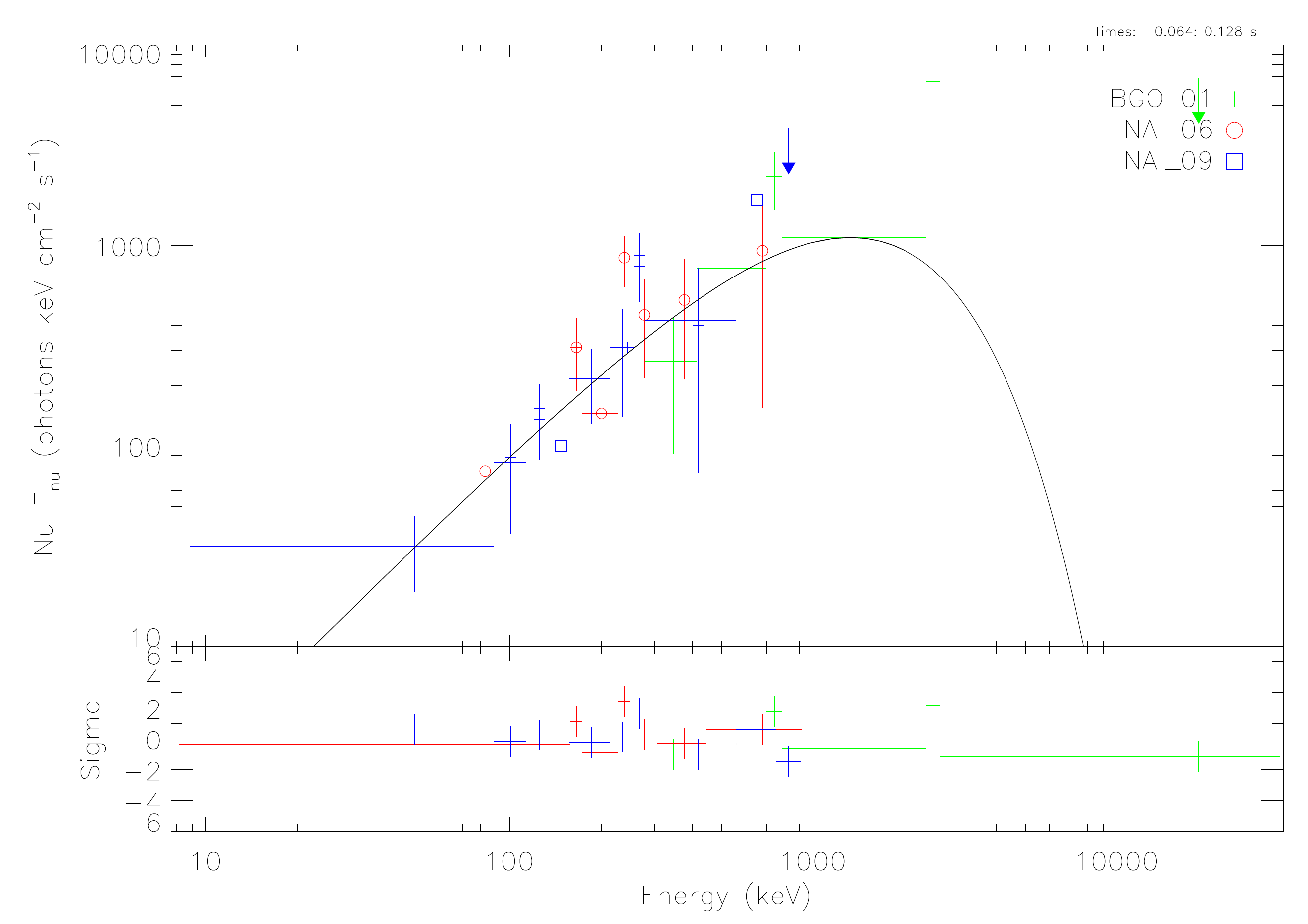} 
\caption{The same as in Fig.~\ref{aba:fig1}, in the $\Delta T_1$ time interval. A comparison between BB (left panel) and Compt (right panel) models.}
\label{aba:fig2}
\end{figure*}

The spectrum of the $\Delta T_1$ time interval, which can be interpreted as the P-GRB emission, is equally best-fit, among all the possible models, by a black body (BB) and a Compt spectral models.
Fig.~\ref{aba:fig2} and Table~\ref{tab:1} illustrate the results of this time-resolved analysis.
From the difference in the C-STAT values between the BB and the Compt models ($\Delta$C-STAT$=9.88$, see Tab.~\ref{tab:1}), we conclude that the simpler BB model can be excluded at $>3\sigma$ confidence level. 
Therefore the best fit is the Compt model.

As in the case of GRB 090510, a Compt spectrum for the P-GRB emission can be interpreted as the result of the convolution of BB spectra at different Doppler factors arising from the a spinning BH \citep[see Sec.~\ref{sec:3} and][]{2016ApJ...831..178R}.

\begin{figure*}
\centering
\includegraphics[height=0.49\hsize,clip,angle=90]{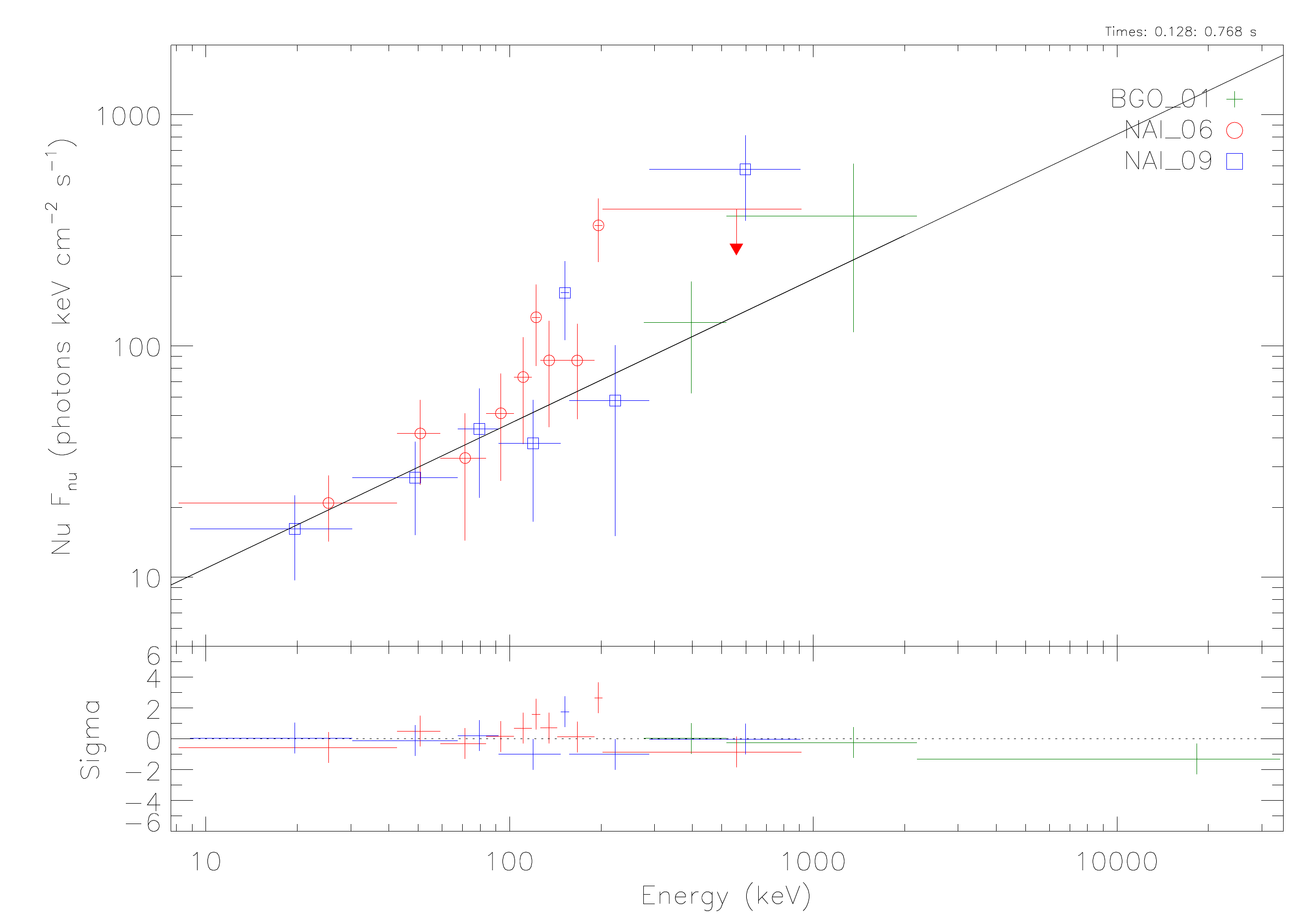}
\includegraphics[height=0.49\hsize,clip,angle=90]{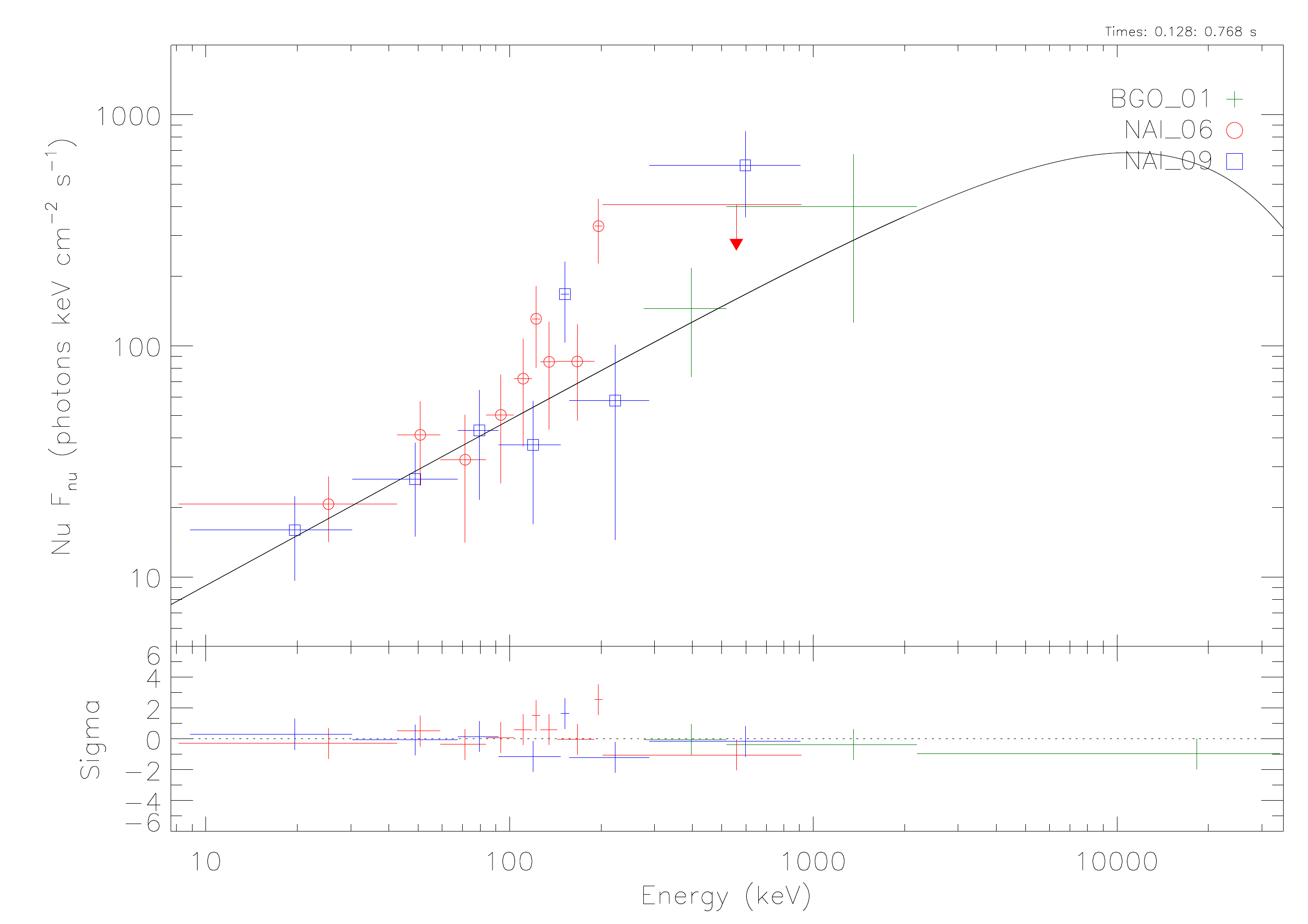} 
\caption{The same as in Fig.~\ref{aba:fig2}, in the $\Delta T_2$ time interval. A comparison between PL (left panel) and Compt (right panel) models.}
\label{aba:fig3}
\end{figure*}

The spectrum of the $\Delta T_2$ time interval, which can be interpretated as the prompt emission, is equally best-fit by a power-law (PL) and a Compt spectral models (see Fig.~\ref{aba:fig3} and Table~\ref{tab:1}).
The PL and the Compt models are equivalent, though Compt model slightly improves the C-STAT statistic.
However, because of the unconstrained value for the peak energy of the Compt model $E_p$, we conclude that the PL model represents an acceptable fit to the data.

%%%%%%%%%%%%%%%%%%%%%%%%%%%%%%%%%%%%%%%%%%%%%%%%%%%%%%
\subsection{Theoretical interpretation within the fireshell model}\label{sec:2.2}
%%%%%%%%%%%%%%%%%%%%%%%%%%%%%%%%%%%%%%%%%%%%%%%%%%%%%%

We proceed to the interpretation of the data analysis performed in Sec.~\ref{sec:2.1} within the fireshell model. 

\subsubsection{The estimate of the redshift}\label{sec:2.2.1}

The identification of the P-GRB and of the prompt emission is fundamental in order to estimate the source cosmological redshift and, consequently, to determine all the physical properties of the $e^+e^-$ plasma at the transparency point \citep{Muccino2012,2015ApJ...808..190R}.
The method introduced in \citet{Muccino2012} allows to determine the source redshift from two main observational constraints: the observed P-GRB temperature $kT$, related to the theoretically-computed rest-frame temperature $kT_{blue}=kT(1+z)$, and the ratio between the P-GRB fluence $S_{BB}=F(\Delta T_1)\Delta T_1$ and the total one $S_{tot}=F(T_{90})T_{90}$, which represents a good redshift independent approximation for the ratio $E_{\rm P-GRB}/E_{e^+e^-}$ (see Tab.~\ref{tab:1}). 
A trial and error procedure is then started, using various set of values for $E_{e^+e^-}^{\rm tot}$ and $B$ to reproduce the observational constraints. 
Each of these set of values provides various possible values for the redshift $z$ from the relation between $kT$ and $kT_{blue}$. 
The closure condition is represented by the $E_{\rm iso}(z)\equiv E_{e^+e^-}^{tot}$, where $E_{\rm iso}$ is computed taking into account the $K$-correction on $S_{tot}$ \citep{Schaefer2007}. 
The redshift verifying the last condition and the corresponding values of $E_{e^+e^-}^{\rm tot}$ and $B$ are the correct one for the source.   
The theoretical redshift $z=3.12\pm1.82$ together with all the other quantities so far determined are summarized in Tab.~\ref{tab:2} (for further details on the method see, e.g., \citealp{2015ApJ...808..190R}).
The analogy with the prototypical source GRB 090227B ($B=4.13\times10^{-5}$, \citealp{Muccino2012}), GRB 140619B ($B=5.52\times10^{-5}$, \citealp{2015ApJ...808..190R}), and GRB 090510 ($B=5.54\times10^{-5}$, \citealt{2016ApJ...831..178R}) is very striking.

The self-consistency of the above theoretical method to estimate the redshift has been tested in S-GRB 090510 \citep{2016ApJ...831..178R}. In this case a theoretical redshift $z_{\rm th}=0.75\pm0.17$ has been derived, in agreement with the spectroscopic measurement $z=0.903\pm0.003$ \citep{GCN9353}.

\subsubsection{Analysis of the prompt emission}\label{sec:2.2.2}

\begin{figure}
\centering
\includegraphics[width=0.9\hsize,clip]{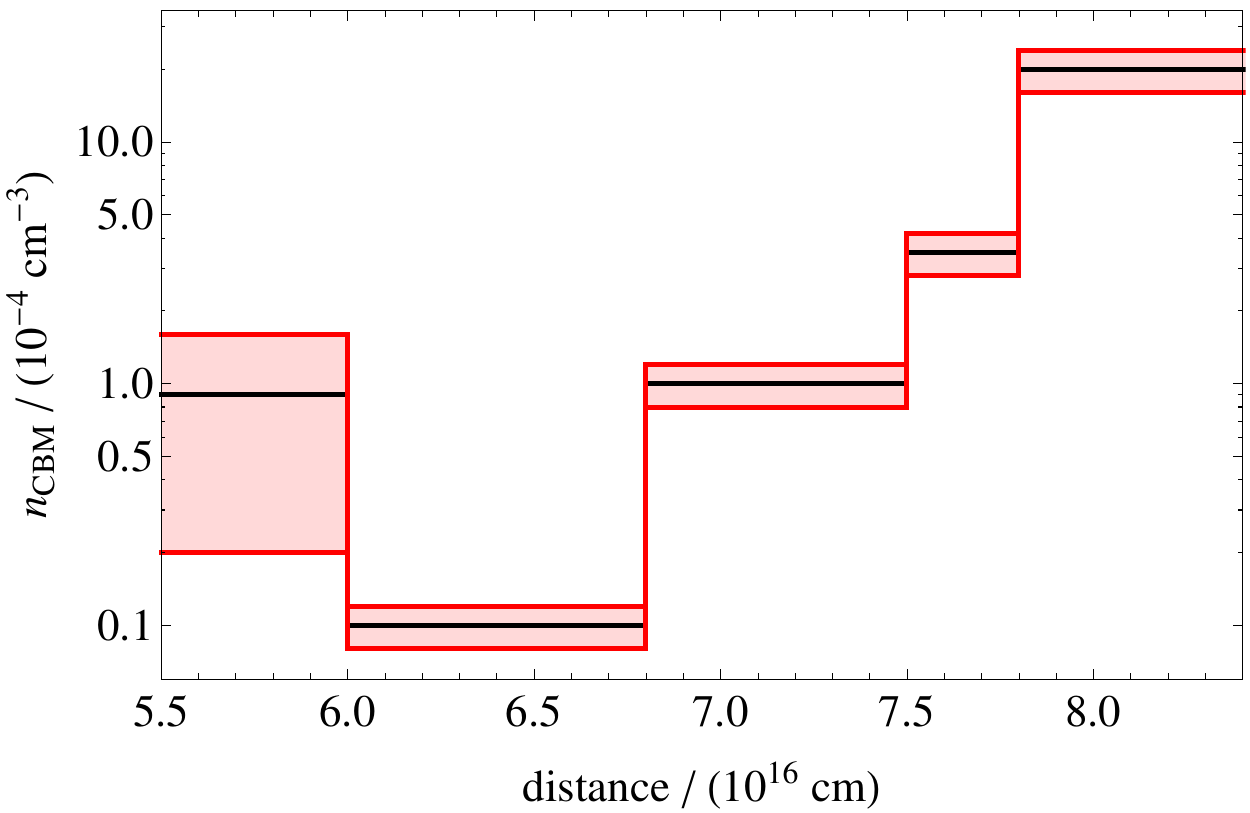}\\
\includegraphics[width=0.9\hsize,clip]{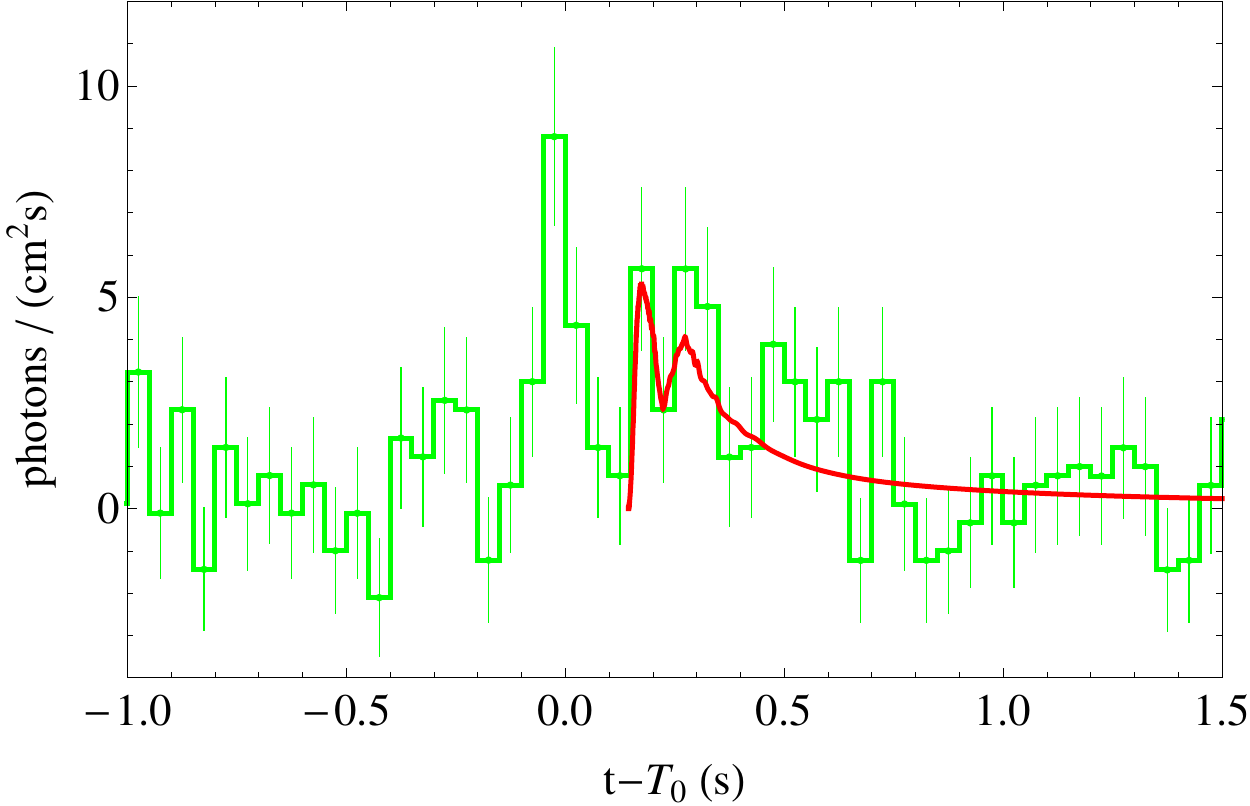}\\
\includegraphics[width=0.9\hsize,clip]{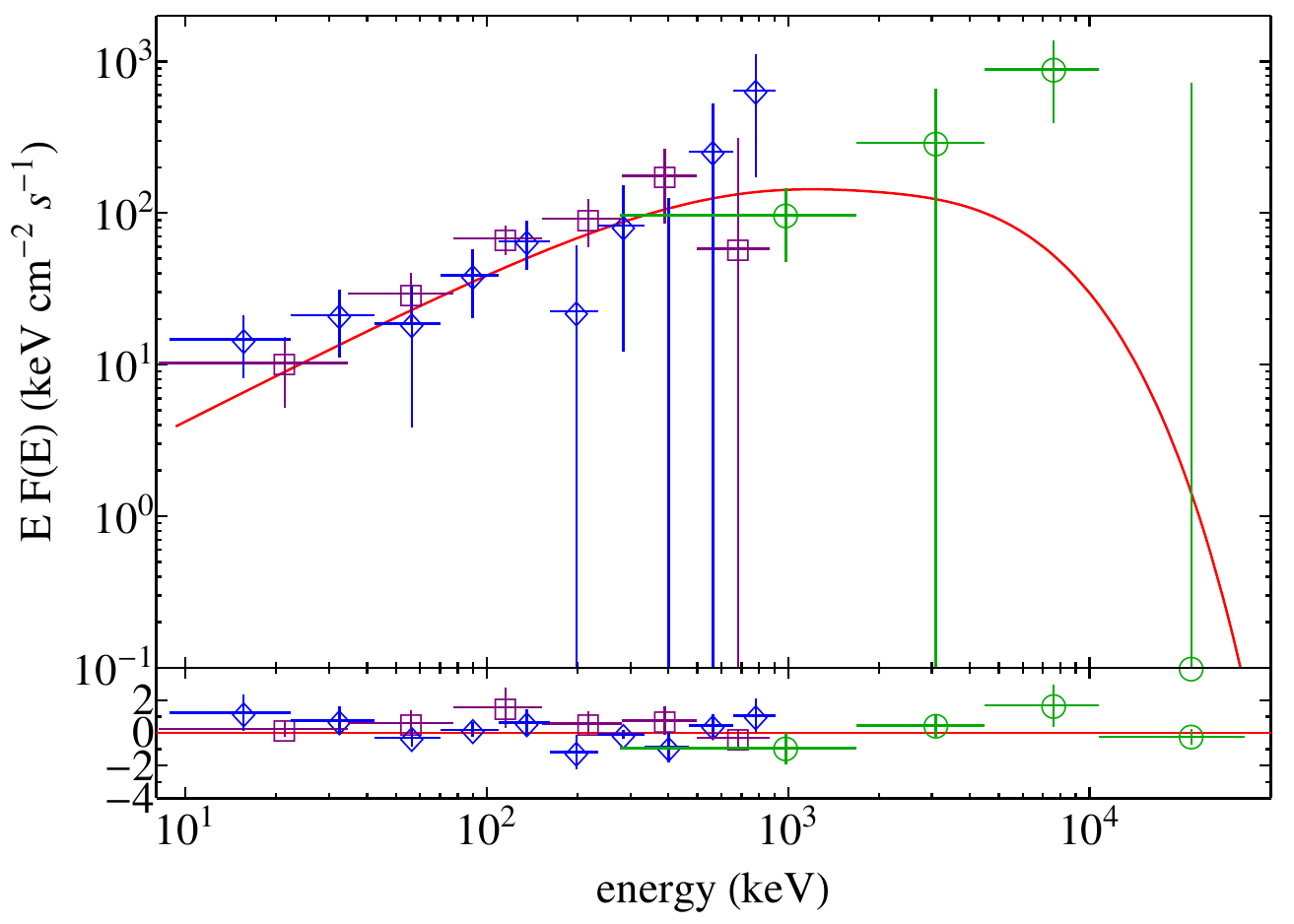}
\caption{Results of the prompt emission simulation of the S-GRB 081024B. Top: the CBM number density (black line) and errors (red shaded region). Middle: comparison between the simulated prompt emission light curve (solid red curves) and the NaI-n9 ($8$ -- $900$ keV) data. Bottom: comparison between the simulated spectrum (solid red curve) and the NaI-n6 (purple squares), NaI-n9 (blue diamonds), and the BGO-b1 (green circles) spectra within the $\Delta T_2$ time interval. The residuals are shown in the sub-plot.}
\label{fig:6}
\end{figure}

In the fireshell model, the prompt emission light curve is the result of the interaction of the accelerated baryons with the CBM \citep[see above and, e.g.,][]{Ruffini2002,Ruffini2006,Patricelli}.
After the determination of the initial conditions for the fireshell, i.e., $E^{tot}_{e^+e^-}$ and $B$ (see Tab.~\ref{tab:2}), to simulate the prompt emission light curve of the S-GRB 081024B (see Figs.~\ref{fig:f1a}) and its corresponding spectrum, we derived the CBM number density and the filling factor $\mathcal{R}$ distributions and the corresponding attached errors (see Tab.~\ref{tab:2} and Fig.~\ref{fig:6}, top panel).
The average CBM number density inferred from the prompt emissions of GRB 081024B is $\langle n_{\rm CBM} \rangle=(3.18\pm0.74)\times10^{-4}$ (see Tab.~\ref{tab:2}), and is larger than those of GRB 140619B, $\langle n_{CBM} \rangle = (4.7\pm1.2)\times10^{-5}$ cm$^{-3}$ \citep{2015ApJ...808..190R}, and GRB 090227B, $\langle n_{CBM} \rangle = (1.90\pm0.20)\times10^{-5}$ cm$^{-3}$ \citep{Muccino2012}, but still typical of the S-GRB galactic halo environments.

\begin{table*}
\scriptsize
\centering
\begin{tabular}{cccccccc}
\hline\hline
\multicolumn{8}{c}{P-GRB}\\
\hline
$z$       &  $E^{tot}_{e^+e^-}/$($10^{52}$~erg)        &  $B/10^{-5}$       & $M_{\rm B}/$($10^{-7}$~M$_\odot$)   &  $E_{\rm P-GRB}/E_{e^+e^-}$ ($\%$)  &  $\Gamma_{\rm tr}/10^4$                          
                &   $r_{\rm tr}/$($10^{12}$~cm)                          &  $kT_{blue}$ (MeV)         \\
\hline
 $3.12\pm1.82$    &   $2.64\pm1.00$      &  $4.6\pm2.8$  &  $6.7\pm4.8$  &  $50\pm26$                                       &  $1.10\pm0.24$  
                & $5.6\pm2.1$   &  $1.39\pm0.76$	                                                  \\
\hline\hline
\multicolumn{8}{c}{Prompt}\\
\hline
Cloud        &  $r$ (cm)  & $\Delta r$ (cm)            &  $n_{CBM}/$($10^{-4}$~cm$^{-3}$)         &       $M_{\rm CBM}/$($10^{22}$~g)     &  $\mathcal{R}/10^{-12}$                       &  $\Gamma/10^4$            &  $d_{\rm v}$ (cm)               \\
\hline
$1^{st}$   &  $5.5\times10^{16}$  &  $0.5\times10^{16}$  &  $0.90\pm0.70$    &  $3.1\pm2.4$ &  $9.0\pm7.0$  &  
$1.10$   &  $2.90\times10^{10}$\\
 $2^{nd}$   &  $6.0\times10^{16}$  & $0.8\times10^{16}$ & $0.10\pm0.02$    & $0.69\pm0.14$   & &  
$0.52$   &  $6.60\times10^{14}$\\
 $3^{rd}$   &  $6.8\times10^{16}$  & $0.7\times10^{16}$ & $1.00\pm0.20$    & $7.5\pm1.5$  &  &  
$0.51$   &  $7.68\times10^{14}$\\
 $4^{th}$   &  $7.5\times10^{16}$  & $0.3\times10^{16}$ & $3.5\pm0.70$   & $12.9\pm2.6$ &  $98\pm53$  &  
$0.40$   &  $1.08\times10^{15}$\\
 $5^{th}$   &  $7.8\times10^{16}$  & $0.7\times10^{16}$ & $20.0\pm4.00$    &  $196\pm39$ &   &  
$0.29$   &  $1.55\times10^{15}$\\
\hline
 average   &   &   &  $3.18\pm0.74$    &   &   &   &   \\
\hline
\end{tabular}
\caption{The P-GRB and prompt emission parameters of the S-GRBs 081024B within the fireshell model.
The P-GRB list of parameters (upper part of the table) are: the inferred redshift $z$, the $e^+e^-$ plasma energy $E^{tot}_{e^+e^-}$, the baryon load $B$ and the corresponding baryonic mass $M_{\rm B}$, the P-GRB energy $E_{\rm P-GRB}$ over $E^{tot}_{e^+e^-}$, and the Lorentz factor $\Gamma_{\rm tr}$, the radius of the fireshell $r_{\rm tr}$, and the temperature blue-shifted toward the observed $kT_{blue}$ computed at the transparency point.
The CBM properties inferred from the prompt emission simulation (lower part of the table) are: the number of CBM clouds, the distance $r$ from the BH, the thickness $\Delta r$, the number density distribution $n_{\rm CBM}$, the total mass $M_{\rm CBM}$, the filling factors $\mathcal{R}$, the Lorentz factor after each collision $\Gamma$, and the total transversal sizes $d_{\rm v}$ of the fireshell visible area. The average number density is indicated at the end of the $n_{\rm CBM}$ column.}
\label{tab:2}
\end{table*}

The simulation of the prompt emission light curve of the NaI-n9 ($8$ -- $900$ keV) data of GRB 081024B is shown in Fig.~\ref{fig:6} (middle panel). 
The short time scale variability observed in the S-GRB light curves is the result of the large values of the Lorentz factor ($\Gamma\approx10^4$, see Tab.~\ref{tab:2}). Under these conditions the total transversal size of the fireshell visible area, $d_v$, is smaller than the thickness of the inhomogeneities ($\approx10^{16}$~cm, see the values indicated in Tab.~\ref{tab:2}), justifying the spherical symmetry approximation \citep{Ruffini2002,Ruffini2006,Patricelli} and explaining the no significant ``broadening'' in arrival time of the luminosity peaks.

The corresponding spectrum is simulated by using the spectral model described in \citet{Patricelli} with phenomenological parameters $\bar{\alpha}=-1.99$.
The rebinned data within the $\Delta T_2$ time interval agree with the simulation, as shown by the residuals around the fireshell simulated spectrum (see Fig.~\ref{fig:6}, bottom panel).

\section{The S-GRB 140402A}\label{sec:4}

\subsection{Observations and data analysis}\label{sec:4.1}

The short hard GRB 140402A was detected on 2014 April 2 at 00:10:07.00 (UT) by the \textit{Fermi}-GBM \citep{2014GCN..16070...1J}.
The duration of this S-GRB in the $50$--$300$ keV is $T_{90}=0.3$~s.
It was also detected by the \textit{Fermi}-LAT \citep{2014GCN..16069...1B} with a best on-ground location $(RA,\,Dec)=(207^{\rm{o}}.47,\,5^{\rm{o}}.87)$ (J2000), consistent with the GBM one.
More than $10$ photons were detected above $100$~MeV and within $10^{\rm{o}}$ from the GBM location, which spatially and temporally correlates with the GBM emission with high significance \citep{2014GCN..16069...1B}.

This burst was also detected by the \textit{Swift}-BAT \citep{2014GCN..16073...1C}, with a best location $(RA,\,Dec)=(207^{\rm{o}}.592,\,5^{\rm{o}}.971)$ (J2000).
No source was detected in the \textit{Swift}-XRT data \citep{2014GCN..16078...1P} after two pointings in PC mode, from $33.3$~ks to $51.2$~ks and from $56$~ks to $107$~ks, respectively. These two observation set are within the $3$-sigma upper limit of the count rate of $3.6\times10^{-3}$~counts/s and $3.0\times10^{-3}$~counts/s, respectively \citep{2014GCN..16078...1P}.
Optical exposures at the full refined BAT position \citep{2014GCN..16073...1C} took by the \textit{Swift}-UVOT (during both the XRT pointings, \citealp{2014GCN..16077...1B}) and by Magellan (at $1.21$ days after the burst, \citealp{2014GCN..16080...1F}) showed no optical afterglow. This allowed to set, respectively, $3$-sigma upper limits of $v>19.8$~mag and of $r>25.0$~mag. 
Consequently, no host galaxy has been associated to this burst and, therefore, no spectroscopic redshift has been determined.

%%%%%%%%%%%%%%%%%%%%%%%%%%%%%%%%%%%%%%%%%%%%%%%%%%%%%%
\subsubsection{Time-integrated spectral analysis of the \textit{Fermi}-GBM data}\label{sec:4.1.1}
%%%%%%%%%%%%%%%%%%%%%%%%%%%%%%%%%%%%%%%%%%%%%%%%%%%%%%

\begin{figure}
\centering
\includegraphics[width=\hsize,clip]{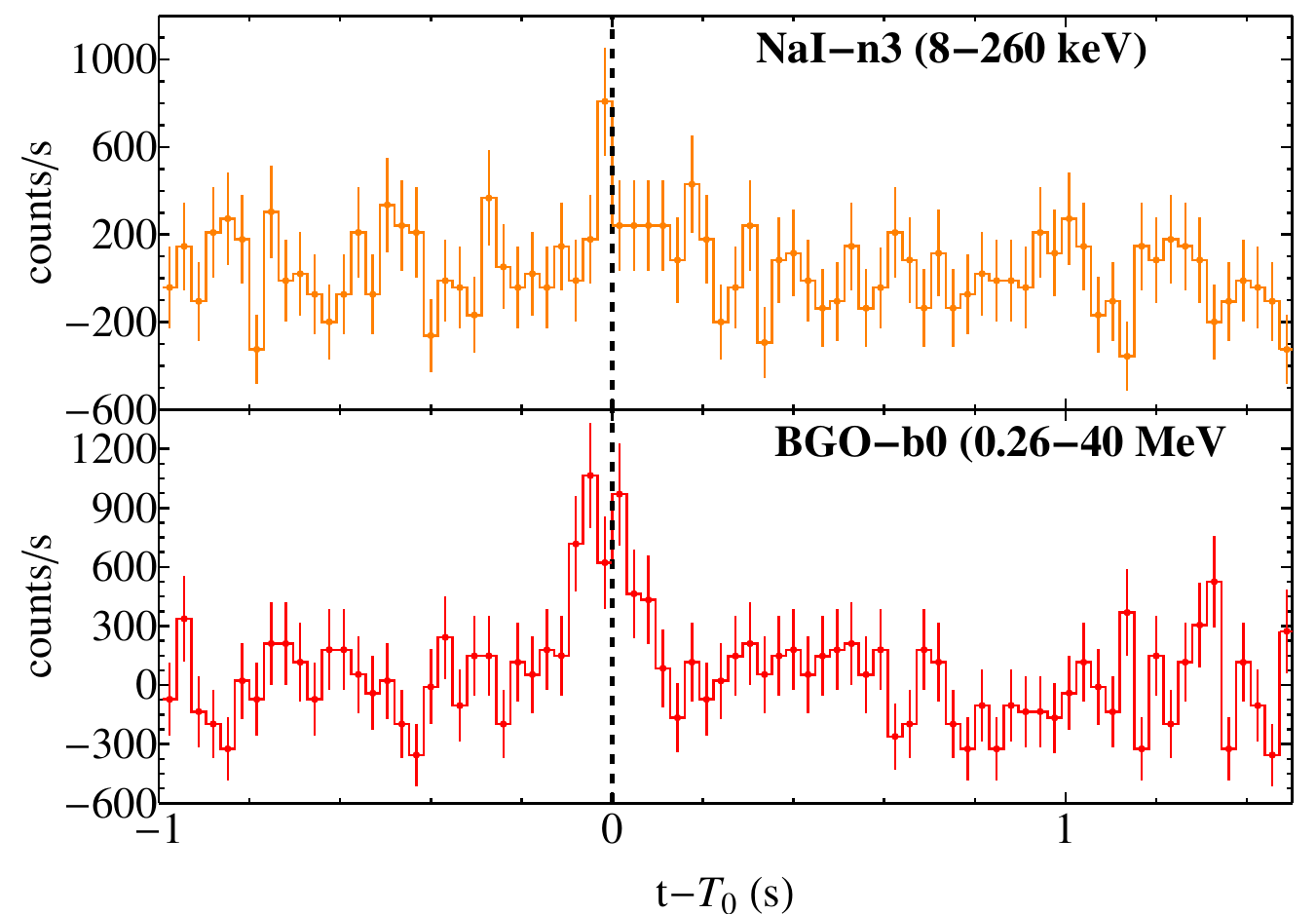}
\caption{Background subtracted light curves of GRB 140402A: the $16$~ms binned light curves from the NaI-n3 ($8$ -- $260$~keV, upper panel) and BGO-b0 ($0.26$ -- $20$~MeV, lower panel) detectors. The vertical dashed line marks the on-set of the LAT light curve (see Fig.~\ref{fig:2}).}
\label{fig:1}
\end{figure}

In Fig.~\ref{fig:1} we reproduced the $16$~ms binned GBM light curves corresponding to detectors NaI-n3 ($8$ -- $260$~keV, top panel) and BGO-b0 ($0.26$ -- $20$~MeV, second panel), and the $0.2$ s binned high-energy light curve ($0.1$ -- $100$~GeV, bottom panel). 
Also for this burst all the light curves are background subtracted.

\begin{figure}
\centering
\includegraphics[width=\hsize,clip]{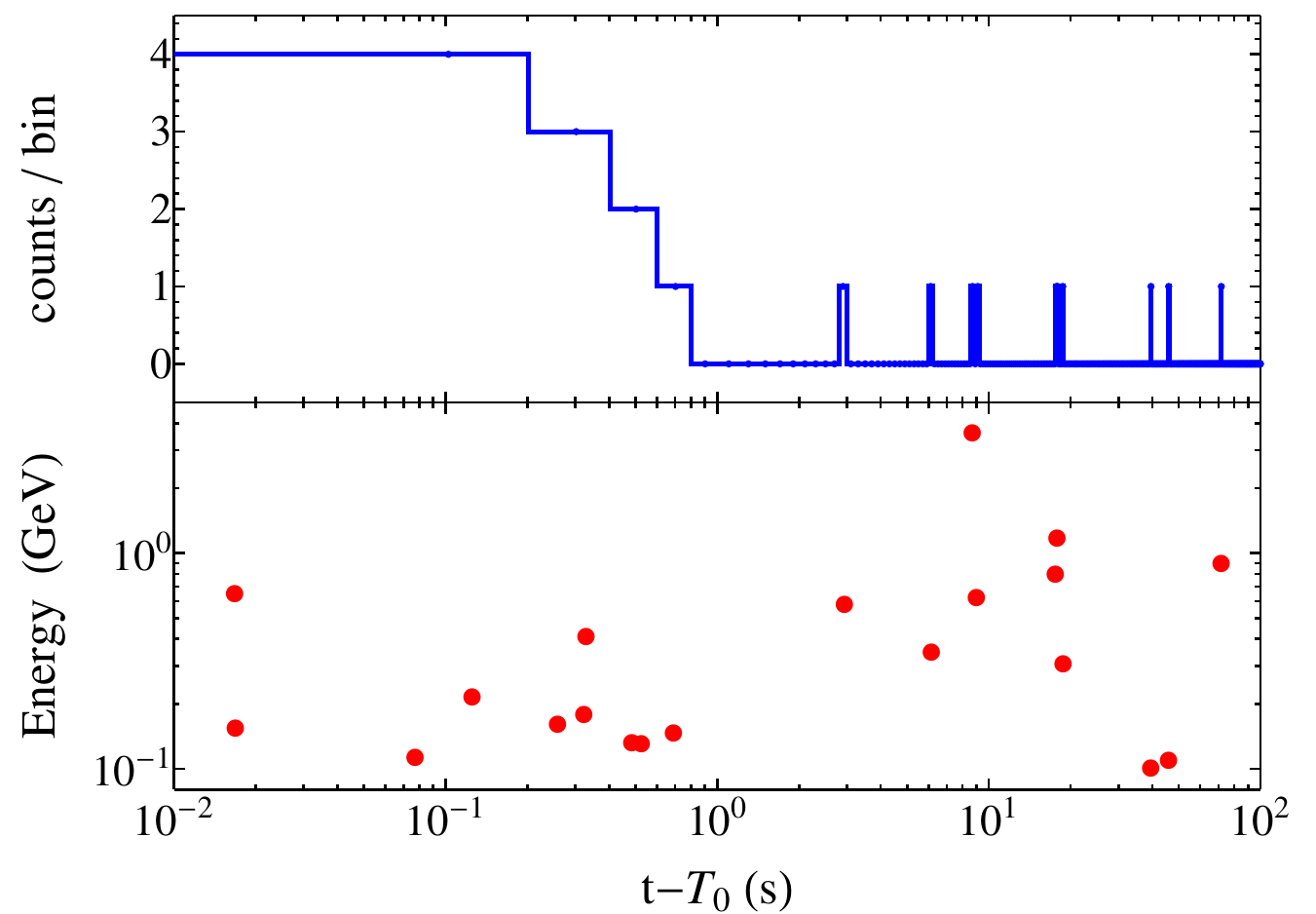} 
\caption{Upper panel: the background subtracted $200$~ms binned high-energy ($0.1$ -- $100$~GeV) light curve, without error bars. Lower panel: the energy light curve of the detected high energy photons.}
\label{fig:2}
\end{figure}

The NaI light curve shows a very weak and short pulse, almost at the background level, while the BGO signal exhibit two sub-structures with a total duration of $\approx0.3$~s. The vertical dashed line in Fig.~\ref{fig:1} represents the on-set of the LAT emission, soon after the first pulse seen in both the GBM light curves.
The background subtracted LAT light curve within $100$~s after the GBM trigger and the corresponding $20$ photons with energies higher than $0.1$~GeV are shown in Fig.~\ref{fig:2}.

\begin{figure*}
\centering
\includegraphics[width=0.49\hsize,clip]{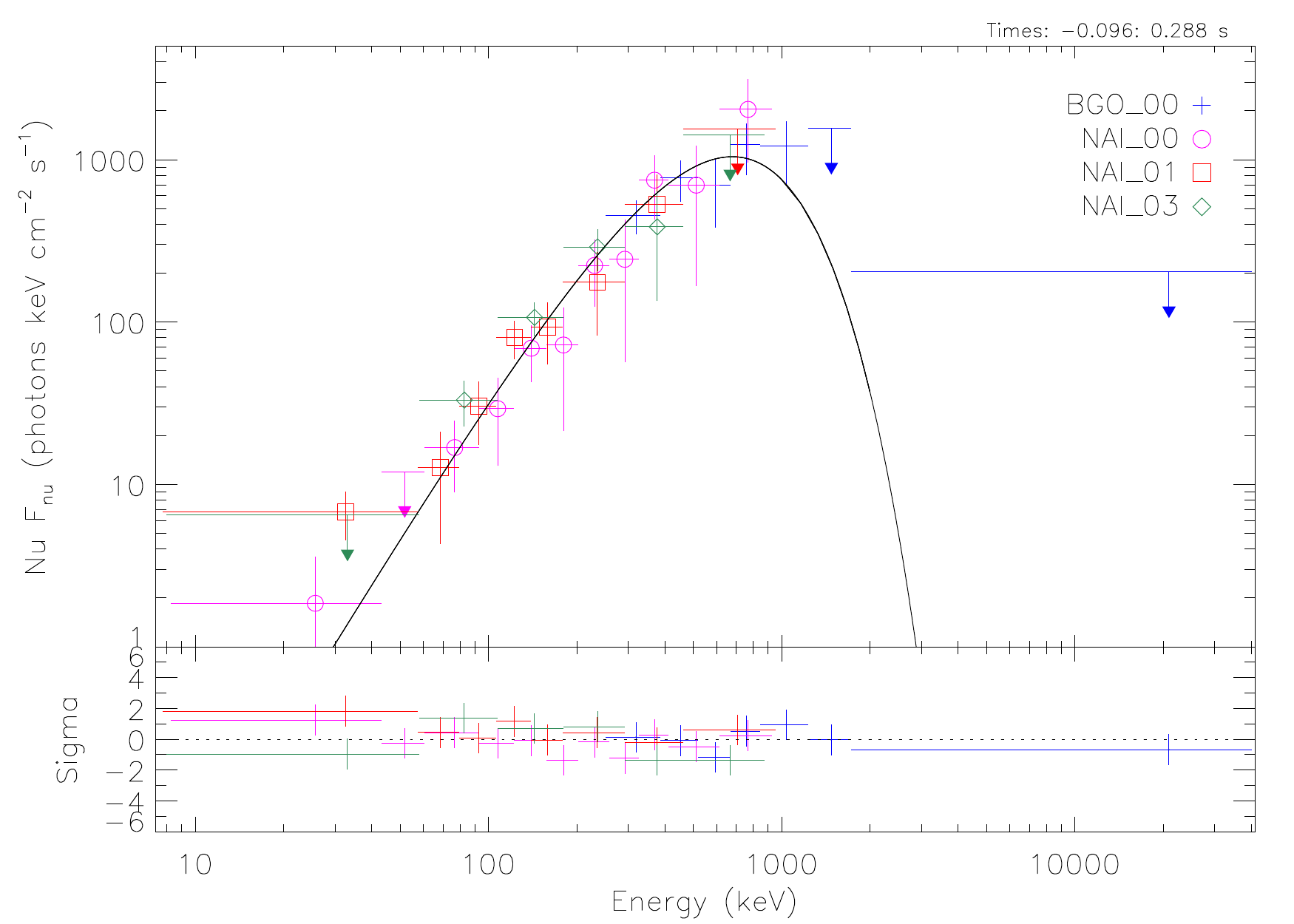}
\includegraphics[width=0.49\hsize,clip]{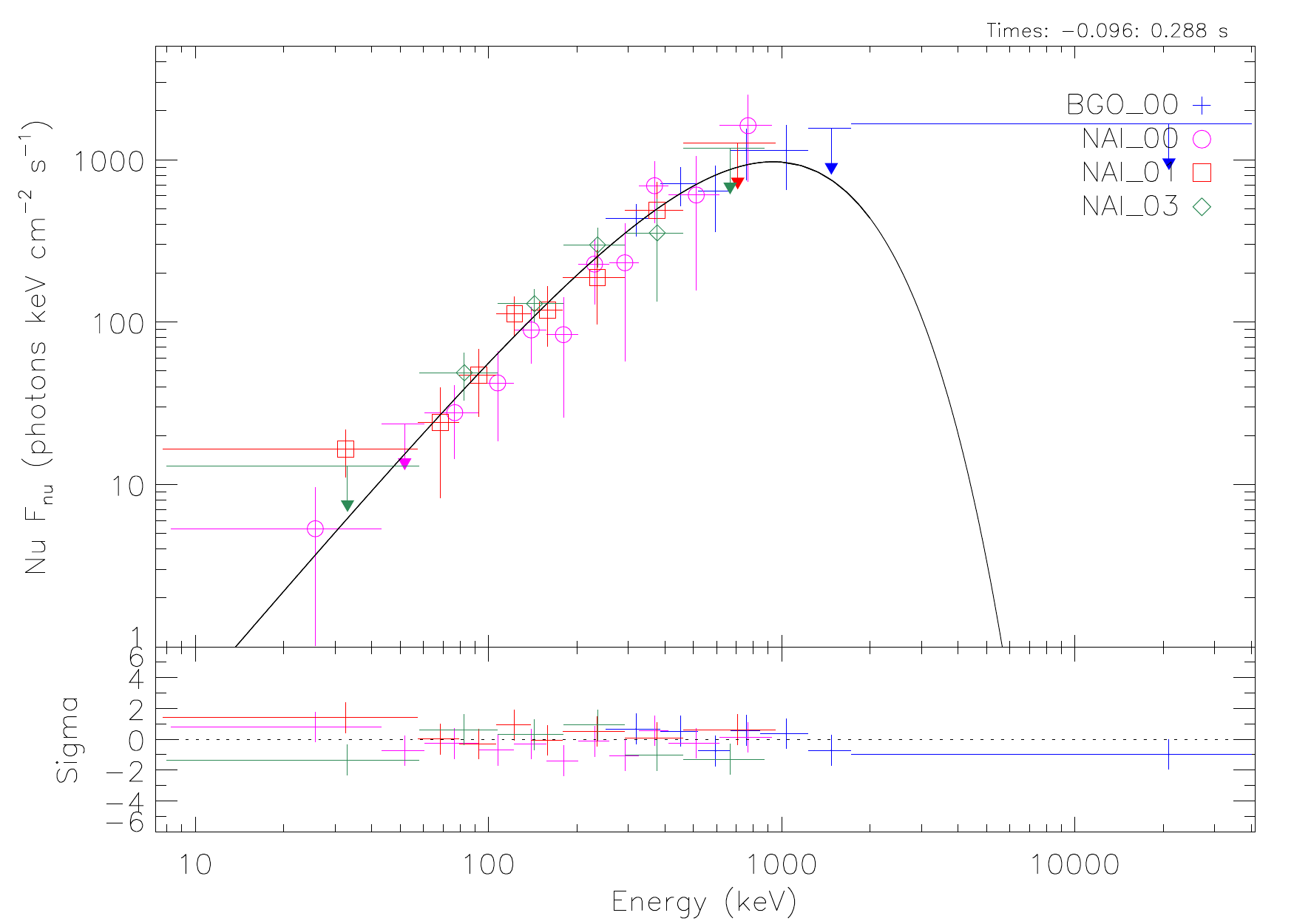}
\caption{The BB (left plot) and the Compt (right plot) spectral fits on the combined NaI--n0, n1, n3+BGO--b1 $\nu F_\nu$ data of GRB 140402A in the $T_{90}$ time interval.}
\label{fig:3}
\end{figure*}

We performed the time-integrated spectral analysis in the time interval from $T_0-0.096$~s to $T_0+0.288$~s (hereafter $T_{90}$). 
To increase the poor statistics at energies $\lesssim260$~keV, we included also the data from the NaI--n0 and n1 detectors in the spectral analysis.
Among all the possible models, BB and Compt equally best-fit the above data  (see Fig.~\ref{fig:3} and the results listed in Tab.~\ref{tab:1a}).
From the value $\Delta$C-STAT$=5.99$ between the above two models (see Tab.~\ref{tab:1a}), we conclude that the Compt model is an acceptable fit to the data.
Similar to the GRB 140619B \citep{2015ApJ...808..190R}, also in the case of GRB 140402A the low-energy index of the Compt model is consistent with $\alpha\sim0$. 
From theoretical and observational considerations on the on-set of the GeV emission (see Sec.~\ref{sec:3} and Fig.~\ref{fig:1}), we investigate the presence of a spectrum consistent with a BB one, which corresponds to the signature of the P-GRB emission for moderately spinning BH \citep[see][]{2016ApJ...831..178R}.

\begin{table*}
\scriptsize
\centering
\begin{tabular}{cccccccccc}
\hline\hline
\multicolumn{9}{c}{S-GRB 140402A}\\
\hline
$\Delta T$          &  Model   &  $K$ (ph keV$^{-1}$ cm$^{-2}$s$^{-1}$)   &  $kT$ (keV)      &  $E_{p}$ (MeV)   &  $\alpha$  &   
                        &  $F$ (erg cm$^{-2}$s$^{-1}$)   &  C-STAT/DOF    \\
\hline
$T_{90}$           &  BB       &  $(2.43\pm0.75)\times10^{-7}$                  &  $173\pm18$    &            &   &               &  $(2.26\pm0.31)\times 10^{-6}$                                                
                        &  $527.65/483$  \\
&                        Compt  &  $(7.0\pm1.4)\times10^{-3}$                     &                       &  $0.94\pm0.24$  &  $0.12\pm0.37$     &  &  $(2.77\pm0.62)\times 10^{-6}$ 
                        &  $521.66/482$  \\
\hline
$\Delta T_{1}$    &  BB       &  $(1.67\pm0.69)\times10^{-7}$                  &  $242\pm34$    &                          &                    &       &  $(6.0\pm1.1)\times 10^{-6}$ 
                        &  $441.01/483$  \\
&                       Compt  &  $(7.1\pm2.6)\times10^{-3}$                      &                      &  $1.20\pm0.32$   &  $0.43\pm0.51$    &  &  $(6.9\pm1.8)\times 10^{-6}$
                        &  $439.61/482$  \\
\hline
$\Delta T_{2}$    &  BB       &  $(5.0\pm2.2)\times10^{-7}$                     &  $122\pm18$    &                          &                &            &  $(1.17\pm0.22)\times10^{-6}$
                        &  $500.42/483$  \\
&                        Compt  &  $(7.5\pm2.0)\times10^{-3}$                     &                       &  $0.70\pm0.25$    &   $0.07\pm0.54$  &   &  $(1.52\pm0.46)\times10^{-6}$           
                        &  $497.57/482$  \\
\hline
\end{tabular}
\caption{Results of the spectral analyses on the S-GRB 140402A. Each column lists: the GRB, the time interval $\Delta T$, the spectral model, the normalization constant $K$, the BB temperature $kT$, the Compt peak energy $E_p$, the low-energy photon index $\alpha$, the $8$ keV -- $40$ MeV energy flux $F$, and the value of the C-STAT over the number of degrees of freedom (DOF).}
\label{tab:1a}
\end{table*}

%%%%%%%%%%%%%%%%%%%%%%%%%%%%%%%%%%%%%%%%%%%%%%%%%%%%%%
\subsubsection{Time-resolved spectral analysis of the \textit{Fermi}-GBM data}\label{sec:4.1.2}
%%%%%%%%%%%%%%%%%%%%%%%%%%%%%%%%%%%%%%%%%%%%%%%%%%%%%%

\begin{figure*}
\centering 
\includegraphics[width=0.49\hsize,clip]{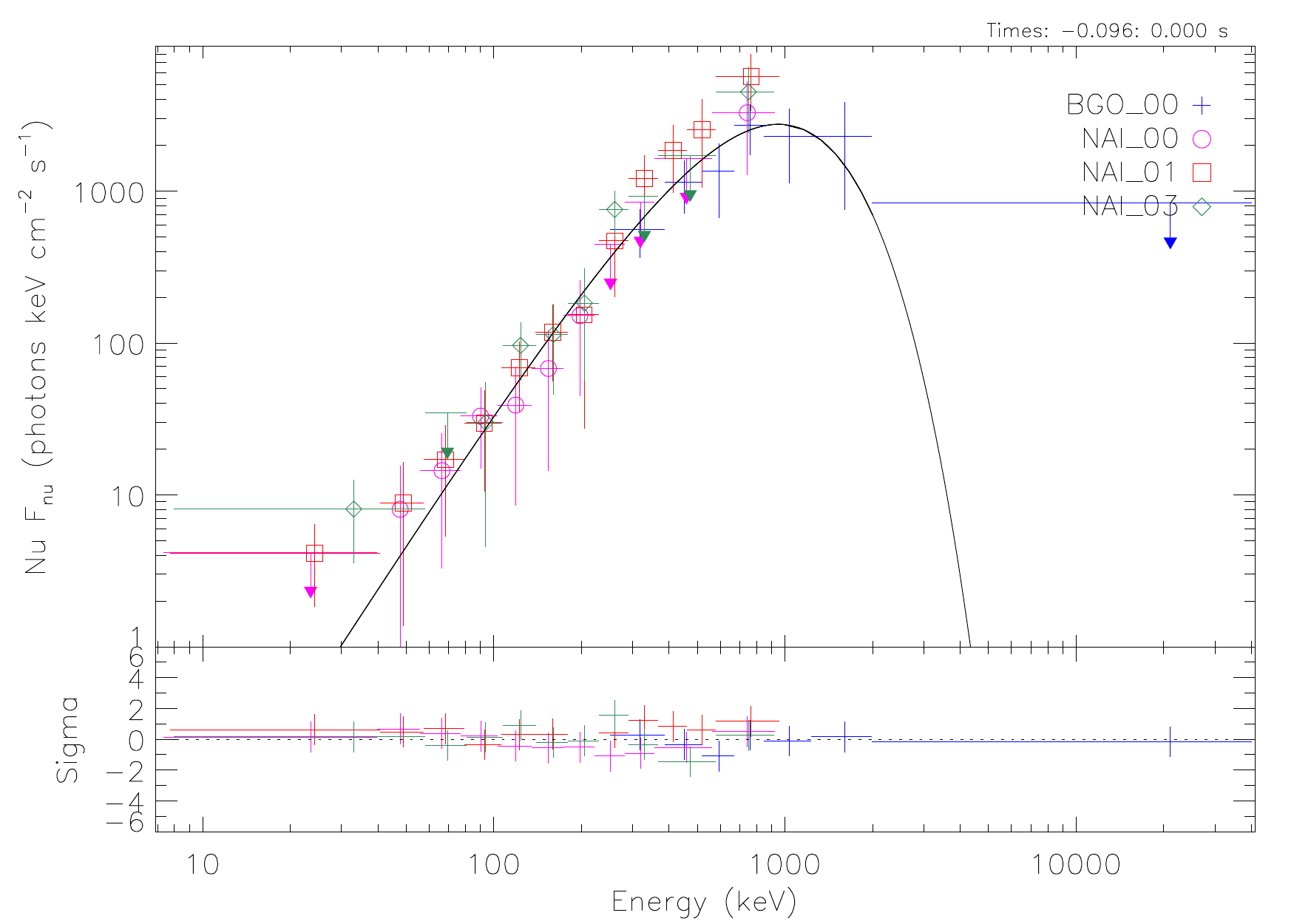}
\includegraphics[width=0.49\hsize,clip]{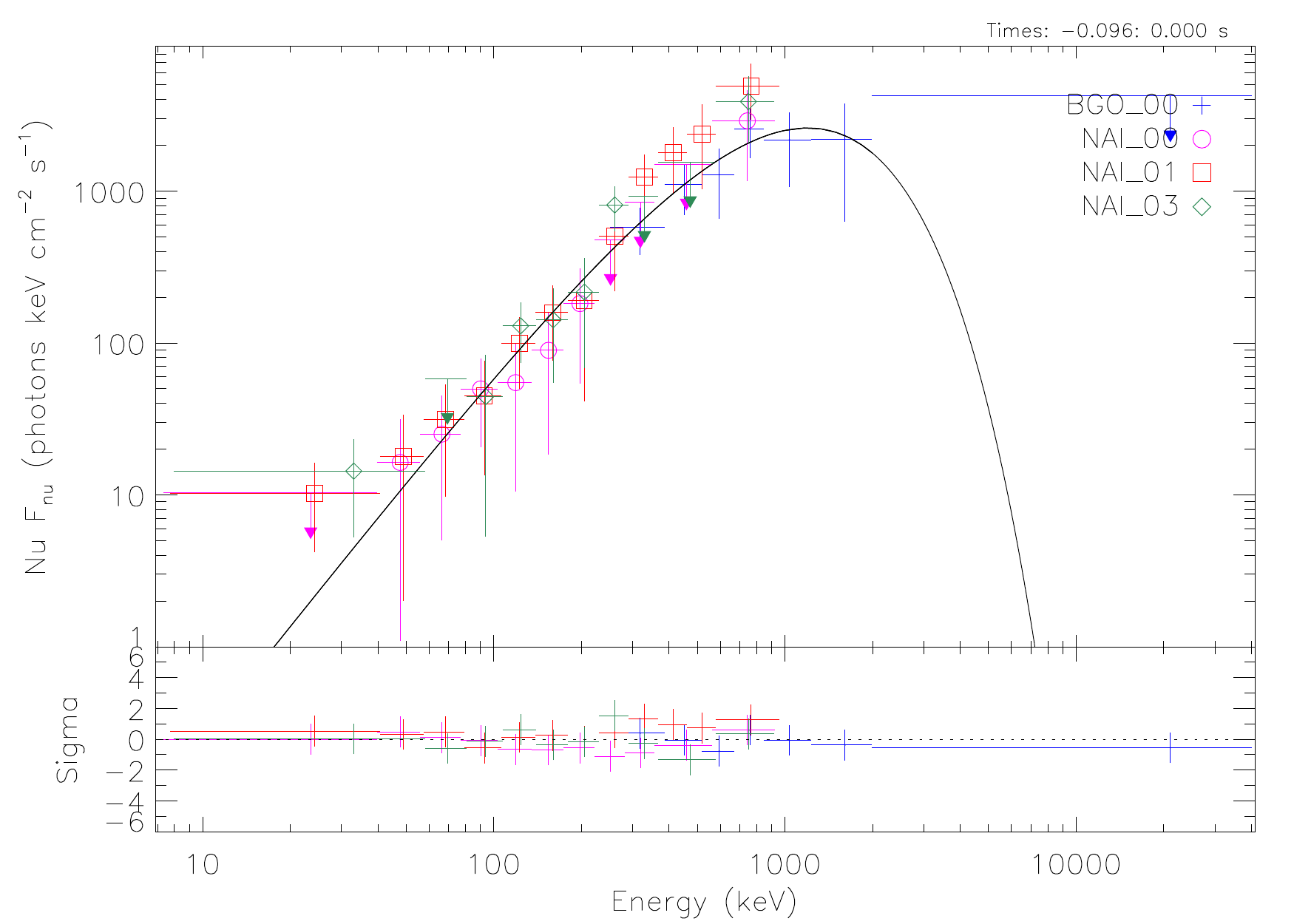} 
\caption{The same as in Fig.~\ref{fig:3}, in the $\Delta T_1$ time interval. A comparison between BB (left panel) and Compt (right panel) models.}
\label{fig:4}
\end{figure*}

The first spike (see Fig.~\ref{fig:1}), observed before the on-set of the GeV, emission extends from $T_0-0.096$~s to $T_0$ (hereafter $\Delta T_1$).
Again BB and Compt spectral models equally best-fit the above data.
As it is shown in Fig.~\ref{fig:4} and Tab.~\ref{tab:1a}, the above two models are almost indistinguishable, with the low-energy index of the Compt model $\alpha=0.43\pm0.51$ being consistent within almost 1-$\sigma$ level with the low energy index of a BB ($\alpha=1$). 
We conclude that the BB model is an acceptable fit to the data and identify the first pulse in the light curve with the P-GRB emission.

\begin{figure*}
\centering 
\includegraphics[width=0.49\hsize,clip]{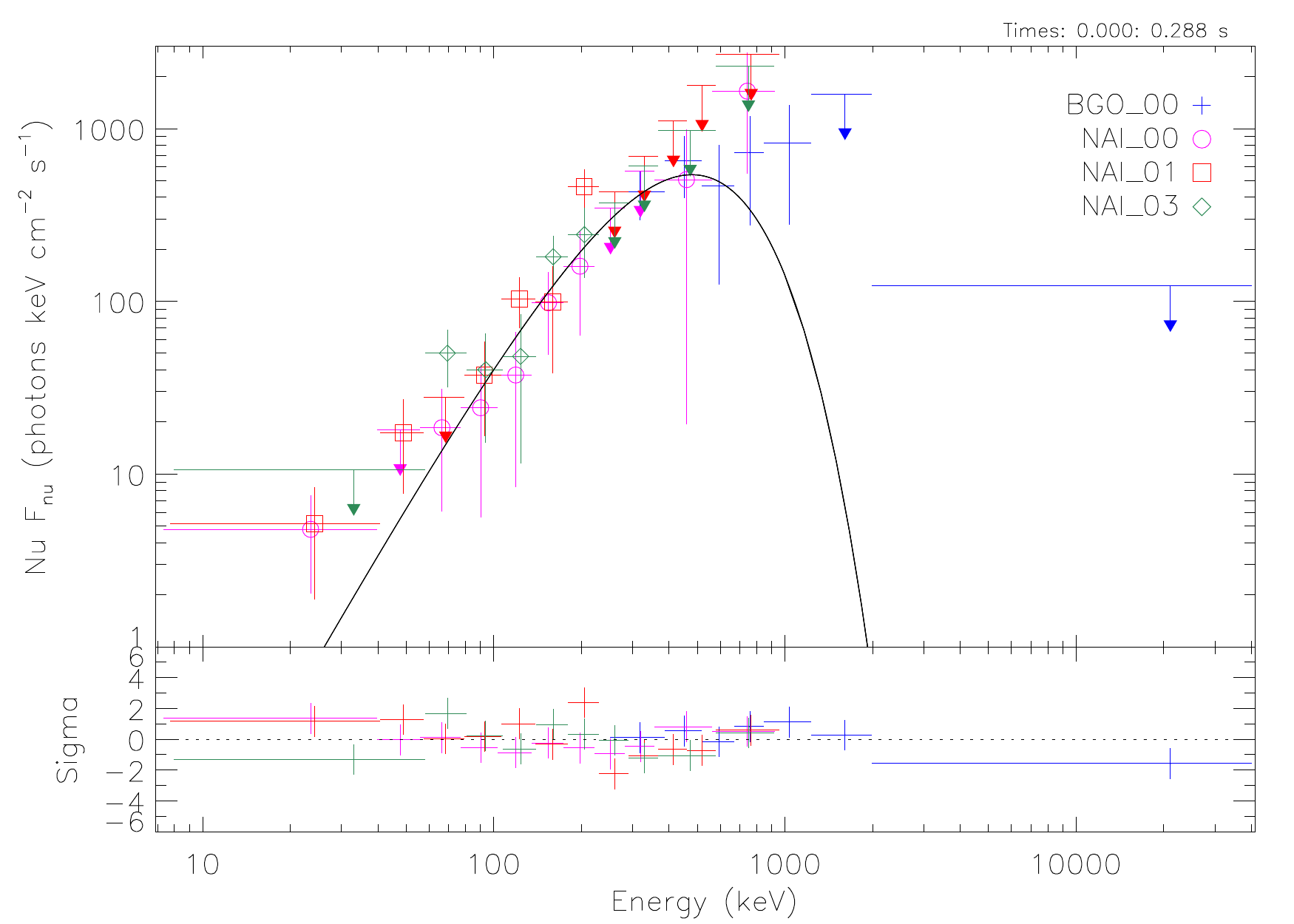}
\includegraphics[width=0.49\hsize,clip]{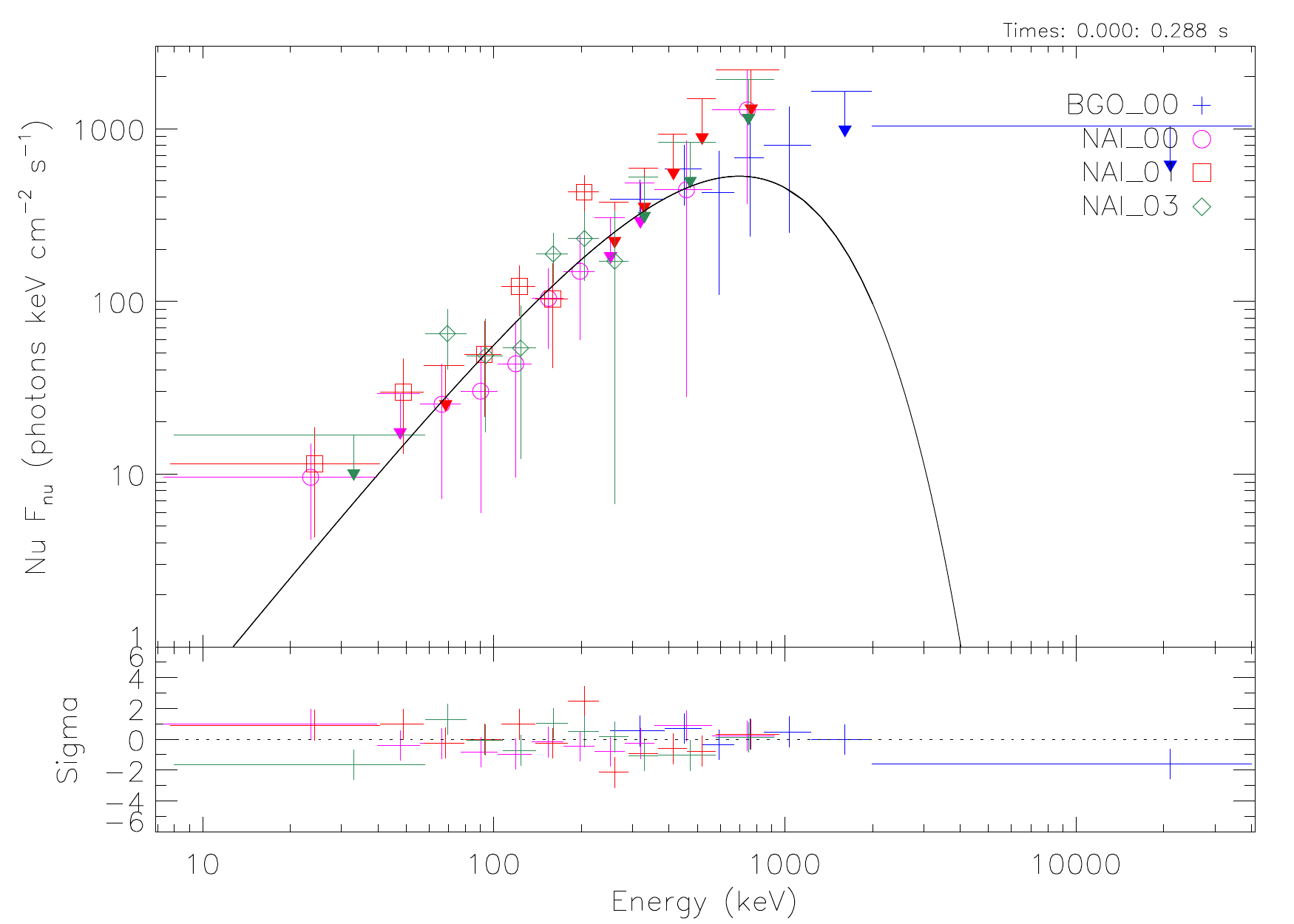} 
\caption{The same as in Fig.~\ref{fig:3}, in the $\Delta T_2$ time interval. A comparison between BB (left panel) and Compt (right panel) models.}
\label{fig:5}
\end{figure*}

The spectrum of the emission in the time interval from $T_0$ to $T_0+0.288$~s (hereafter $\Delta T_2$) reveals that a Compt model fits slightly better the data points at $\approx1$~MeV and its low-energy index $\alpha=0.07\pm0.54$ indicates that the energy distribution is somehow broader than that of a BB model (see Fig.~\ref{fig:5} and Tab.~\ref{tab:1a}).
The Compt model is consistent with the modified BB spectrum adopted in the fireshell model for the prompt emission \citep{Patricelli}.
Therefore we identify the $\Delta T_2$ time interval with the prompt emission.

%%%%%%%%%%%%%%%%%%%%%%%%%%%%%%%%%%%%%%%%%%%%%%%%%%%%%%
\subsection{Theoretical interpretation within the fireshell model}\label{sec:4.2}
%%%%%%%%%%%%%%%%%%%%%%%%%%%%%%%%%%%%%%%%%%%%%%%%%%%%%%

We proceed to the interpretation of the data analysis performed in Sec.~\ref{sec:4.1} within the fireshell model.

\subsubsection{The estimate of the redshift}\label{sec:4.2.1}

After having identified of the P-GRB emission of the S-GRB 140402A (see Sec.~\ref{sec:4.1.2}), we follow the same loop procedure recalled in Sec.~\ref{sec:2.2.1} to infer the redshift, $E_{e^+e^-}^{\rm tot}$ and $B$ of the source. 
The results of this method are summarized in Tab.~\ref{tab:2a}. In particular the theoretically derived redshift for this source is $z=5.52\pm0.93$.
Again, the analogy with the S-GRBs 081024B (see Sec.~\ref{sec:2.1.2}), GRB 090227B \citep{Muccino2012}, 140619B \citep{2015ApJ...808..190R}, and 090510 \citep{2016ApJ...831..178R} is very striking.

\subsubsection{Analysis of the prompt emission}\label{sec:4.2.2}

\begin{figure}
\centering
\includegraphics[width=\hsize,clip]{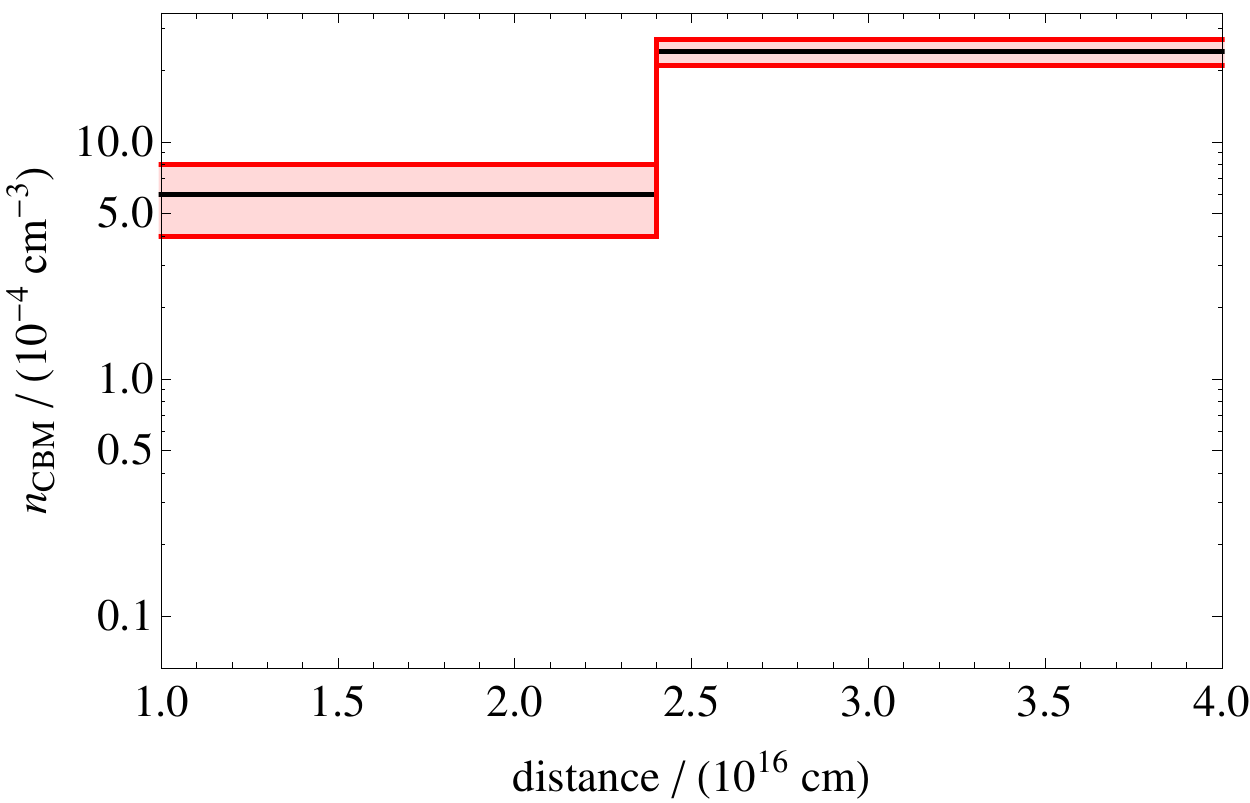}\\
\includegraphics[width=\hsize,clip]{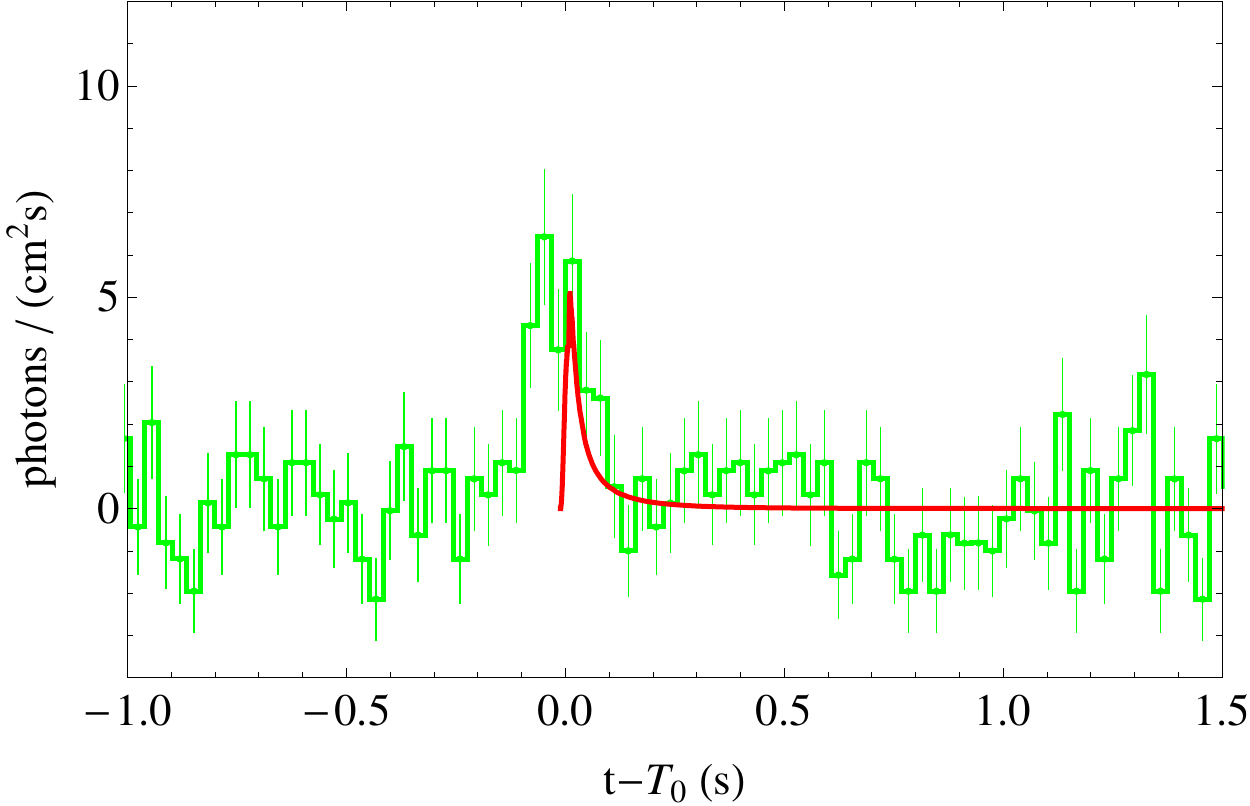}\\
\includegraphics[width=\hsize,clip]{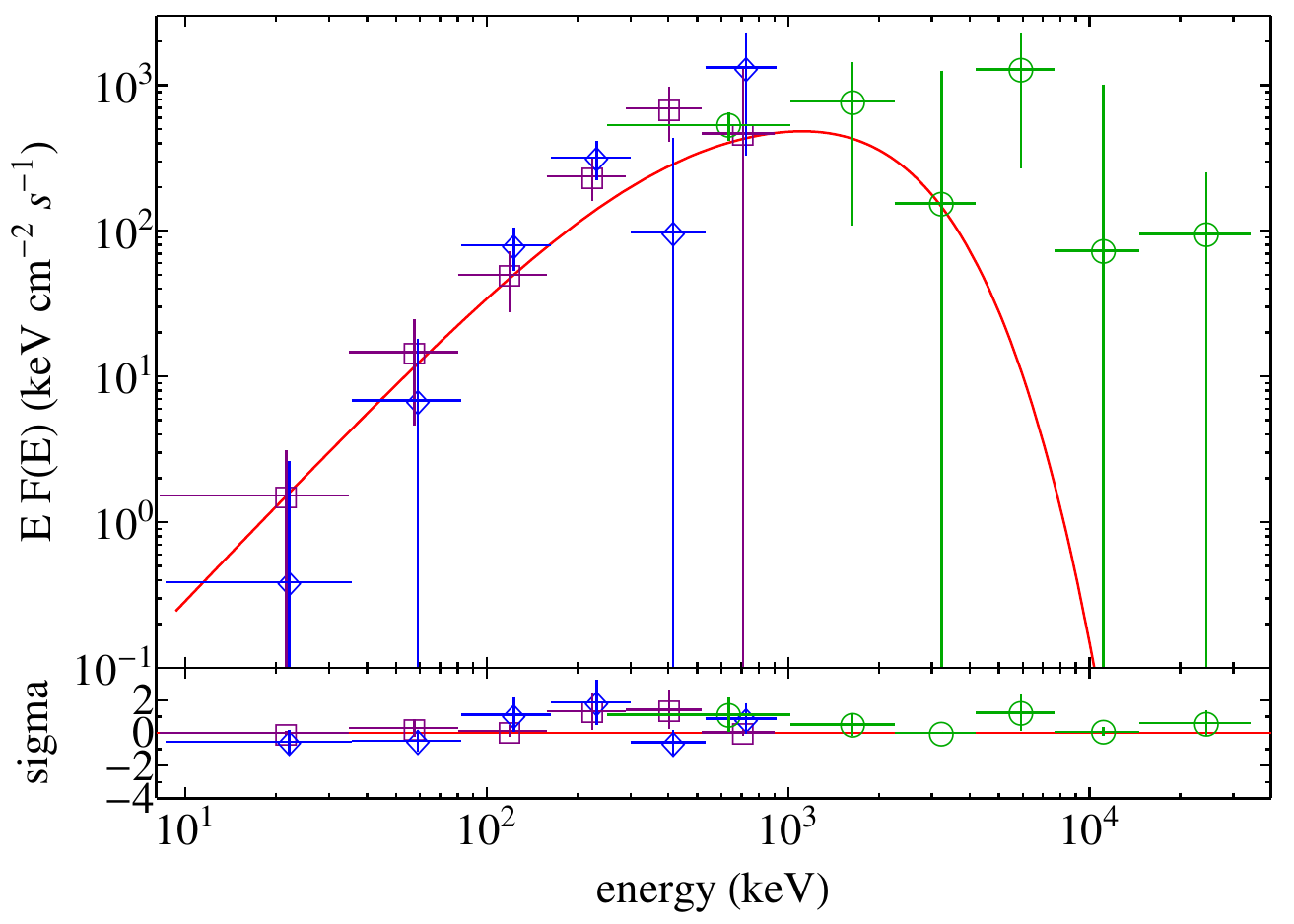}
\caption{Results of the prompt emission simulation of the S-GRB 140402A. Top: the CBM number density (black line) and errors (red shaded region). Middle: comparison between the simulated prompt emission light curve (solid red curves) and the BGO-b0 ($0.26$ -- $40$ MeV) data. Bottom: comparison between the simulated spectrum (solid red curve) and the NaI-n1 (purple squares), NaI-n3 (blue diamonds), and the BGO-b0 (green circles) spectra within the $\Delta T_2$ time interval. The residuals are shown in the sub-plot.}
\label{fig:6a}
\end{figure}

Similarly to the case of the S-GRB 081024B (see Sec.~\ref{sec:2.2.2}), to simulate the prompt emission light curve of the S-GRB 140402A (see Fig.~\ref{fig:1}) and its corresponding spectrum, we derived the CBM number density and the filling factors $\mathcal{R}$ distributions (see Tab.~\ref{tab:2a} and Fig.~\ref{fig:6a}, top panel). 
Also in this case the inferred values fully justify the adopted spherical symmetry approximation \citep{Ruffini2002,Ruffini2006,Patricelli} and explain the negligible ``dispersion'' in arrival time of the luminosity peak. 
\begin{table*}
\scriptsize
\centering
\begin{tabular}{cccccccc}
\hline\hline
\multicolumn{8}{c}{P-GRB}\\
\hline
$z$       &  $E^{tot}_{e^+e^-}/$($10^{52}$~erg)        &  $B/10^{-5}$       & $M_{\rm B}/$($10^{-7}$~M$_\odot$)   &  $E_{\rm P-GRB}/E_{e^+e^-}$ ($\%$) &  $\Gamma_{\rm tr}/10^4$                          
                &   $r_{\rm tr}/$($10^{12}$~cm)                          &  $kT_{blue}$ (MeV)         \\
\hline
$5.52 \pm 0.93$    &   $4.7\pm1.1$      &  $3.6\pm1.0$   &  $9.5\pm3.4$   &  $54\pm16$                                       &  $1.30\pm0.13$  
                &  $6.66\pm0.91$   &  $1.58\pm0.22$                                                  \\
\hline\hline
\multicolumn{8}{c}{Prompt}\\
\hline
Cloud        &  $r$ (cm)  & $\Delta r$ (cm)            &  $n_{CBM}/$($10^{-4}$~cm$^{-3}$)         &       $M_{\rm CBM}/$($10^{22}$~g)     &  $\mathcal{R}/10^{-9}$                       &  $\Gamma/10^4$            &  $d_{\rm v}$ (cm)               \\
\hline
 $1^{st}$   &  $1.0\times10^{16}$  & $1.4\times10^{16}$ &  $6.0\pm2.0$    &  $5.4\pm1.8$  &  $4.7\pm0.45$  &  $1.30$   &  $6.64\times10^{10}$  \\
 $2^{nd}$  &  $2.4\times10^{16}$   & $2.6\times10^{16}$  &  $24.0\pm3.0$   &  $187\pm23$    &                                            &  $0.92$   &  $1.49\times10^{14}$  \\
\hline
 average   &   &   &  $15.4\pm2.5$    &   &   &   &   \\
\hline
\end{tabular}
\caption{The P-GRB and prompt emission parameters of the S-GRBs 140402A within the fireshell model.
For the P-GRB parameters (upper part of the table) and the CBM properties (lower part of the table), inferred from the prompt emission simulation, we refer to Tab.~\ref{tab:2}.}
\label{tab:2a}
\end{table*}

The average CBM number density in the case of GRB 140402A is $\langle n_{\rm CBM} \rangle=(1.54\pm0.25)\times10^{-3}$  (see Tab.~\ref{tab:2a}), which is similar to that inferred from GRB 081024B.
The simulation of the prompt emission light curve of the BGO-b0 ($0.26$ -- $40$ MeV) data of GRB 140402A is shown in Fig.~\ref{fig:6a} (middle panel). 
The simulation of the corresponding spectrum requires a phenomenological parameter  $\bar{\alpha}=-0.9$.
Fig.~\ref{fig:6a} (bottom panel), displays the agreement between the rebinned data from the $\Delta T_2$ time interval with the simulation.

%%%%%%%%%%%%%%%%%%%%%%%%%%%%%%%%%%%%%%%%%%%%%%%%%%%%%%
\section{The GeV emission in S-GRBs}\label{sec:5}
%%%%%%%%%%%%%%%%%%%%%%%%%%%%%%%%%%%%%%%%%%%%%%%%%%%%%%

Before going into more details on the general properties of the S-GRB GeV emission, we briefly summarize the observational features and the data analysis of the high energy emission of the S-GRBs 081024B and 140402A, and then we turn back to a new analysis on the absence of the GeV emission in the S-GRBs 090227B.

\subsection{The GeV emission of the S-GRBs 081024B and 140402A}\label{sec:5.1}

We downloaded the LAT event and spacecraft data\footnote{\url{http://fermi.gsfc.nasa.gov/cgi-bin/ssc/LAT/LATDataQuery.cgi}} selecting the observational time, the energy range and the source coordinates \citep{2014GCN..16069...1B}.
We then made cuts on the dataset time and energy range, position \citep{2014GCN..16069...1B}, region of interest (ROI) radius (typically$10^{\rm{o}}$), and maximum zenith angle.\footnote{The maximum zenith angle selection excludes any portion of the ROI which is too close to the Earth's limb, resulting in elevated background levels.}
Within the event selection recommendations for the analysis of LAT data using the \textit{Pass 8 Data} (P8R2) we adopted the burst and transient analysis (for events lasting $<200$~s) with an energy selection of $0.1$ -- $500$~GeV, a ROI-based zenith angle cut of $100^{\rm{o}}$, an event class $16$, and the instrument response function	\texttt{P8R2\_TRANSIENT020\_V6}.\footnote{\url{http://fermi.gsfc.nasa.gov/ssc/data/analysis/documentation/Cicerone/\\Cicerone\_Data\_Exploration/Data\_preparation.html}}
The additional selection of the good time intervals (GTIs) when the data quality is good (\texttt{DATA\_QUAL>0}) is introduced to exclude time periods when some spacecraft event has affected the quality of the data (in addition to the time selection to the maximum zenith angle cut introduced above).

In the case of the S-GRB 081024B, we obtained the GeV light curve and the observed photon energies showed in Fig.~\ref{fig:f1a} (third and fourth panels), which are in agreement with those reported in \citet{Ackermann2013}.
In the case of the S-GRB 140402A, we obtained the GeV light curve showed in Fig.~\ref{fig:2} (upper plot). 
About $20$ photons with energies higher than $0.1$~GeV have been detected within $100$~s after the GBM trigger (see Fig.~\ref{fig:2}, lower panel).
The highest energy photon is a $3.7$~GeV event which is observed at $T_0+8.7$~s.

Then, we built up the rest-frame $0.1$ -- $100$~GeV light curve of the S-GRBs 081024B and 140402A. 
For the S-GRB 081024B, we rebinned its GeV emission luminosity light curve into two bins, as displayed in \citet{Ackermann2013}.
For the S-GRB 140402A, we rebinned it into two time bins with enough photons to perform a spectral anlysis: from $T_0$ to $T_0+0.6$~s, and from $T_0+0.6$~s to $T_0+20$~s. 

\begin{figure*}
\centering
\includegraphics[width=0.8\hsize,clip]{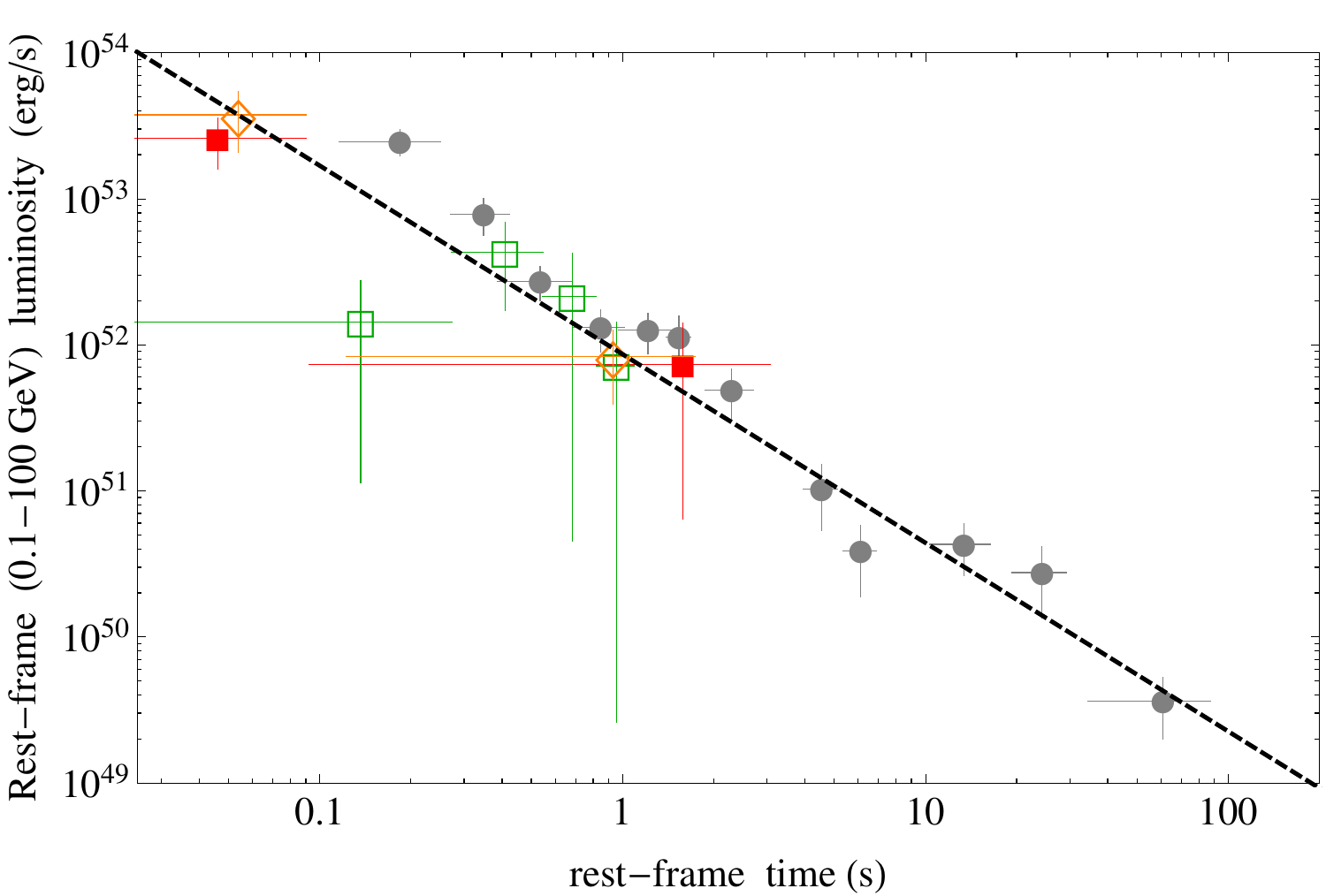}
\caption{The rest-frame $0.1$--$100$ GeV isotropic luminosities of the S-GRBs: 081024B (orange empty diamonds), 090510 (gray filled circles), 140402A (red filled squares), and 140619B (green empty squares). All the light curves are shown from the burst trigger times on, while in the case of the S-GRB 090510 it starts after the precursor emission, i.e., from the P-GRB emission on \citep[see][for details]{2016ApJ...831..178R}. The dashed black line marks the common behavior of all the S-GRB light curves which goes as $t^{-1.29\pm0.06}$.}
\label{fig:7}
\end{figure*}

The resulting luminosity light curves follow a common power-law trend with the rest-frame time which goes as $t^{-1.29\pm0.06}$ (see dashed black line in Fig.~\ref{fig:7}). All the light curves are shown from the burst trigger times on, while in the case of the S-GRB 090510 it starts after the precursor emission, i.e., from the P-GRB emission on \citep[see][for details]{2016ApJ...831..178R}. The GeV emission of the S-GRB 140402A is the second longest in time duration after GRB 090510, which exhibits a common behavior with the light curves of the other S-GRBs after $\sim1$~s rest-frame time (see Fig.~\ref{fig:7}).

Tab.~\ref{tab:LAT} lists the redshift, $E_{p,i}$, $E_{iso}$ (in the rest-frame energy band $1$--$10000$~keV), and the GeV isotropic emission energy $E_{LAT}$ in the rest-frame energy band $0.1$--$100$ GeV of the five authentic S-GRBs discussed here. These values of $E_{LAT}$ are simply obtained by multiplying the average luminosity in each time bin by the corresponding rest-frame duration and, then, by summing up all the contributions for each bin. However, these estimates represent lower limits to the actual GeV isotropic emission energies, since at late times the observations of GeV emission could be prevented due to instrumental threshold of the LAT.   
\begin{table*}
\centering
\begin{tabular}{lcccccccc}
\hline\hline
GRB        &  z                          &  $E_{\rm p,i}$    &    $E_{\rm iso}$      &  $E^{\rm max}_{\rm GeV}$  &  $\Gamma^{\rm min}_{\rm GeV}$  &  $E_{\rm LAT}$      &  $M_{\rm acc}^{\eta_+}$           &  $M_{\rm acc}^{\eta_-}$\\
              &                             &  (MeV)              &    ($10^{52}$~erg)   & (GeV)     &   &  ($10^{52}$~erg)                            & (M$_\odot$)                            &  (M$_\odot$) \\
\hline
081024B  &  $3.12\pm1.82$       &  $9.56\pm4.94$  &  $2.64\pm1.00$        & $3$           &  $\gtrsim779$             &  $\gtrsim2.79\pm0.98$                  &  $\gtrsim0.04$                          &  $\gtrsim0.41$\\
090227B  &  $1.61\pm0.14$       &  $5.89\pm0.30$  &  $28.3\pm1.5$       &    --                        &  --                                    &  --                                               & 
--                                           &  --\\
090510    &  $0.903\pm0.003$    &  $7.89\pm0.76$  &  $3.95\pm0.21$   	    & $31$          &  $\gtrsim551$           &  $\gtrsim5.78\pm0.60$                    &  $\gtrsim0.08$                          &  $\gtrsim0.86$\\
140402A  &  $5.52\pm0.93$       &  $6.1\pm1.6$     &  $4.7\pm1.1$       & $3.7$             &  $\gtrsim354$           &  $\gtrsim4.5\pm2.2$                       &  $\gtrsim0.06$                          &  $\gtrsim0.66$\\
140619B  &  $2.67\pm0.37$       &  $5.34\pm0.79$  &  $6.03\pm0.79$      & $24$         &  $\gtrsim471$                 &   $\gtrsim2.34\pm0.91$                & $\gtrsim0.03$                          &  $\gtrsim0.35$\\
\hline
\end{tabular}
\caption{S-GRB prompt and GeV emission properties. Columns list $z$, $E_{\rm p,i}$, the maximum GeV photon observed energy $E^{\rm max}_{\rm GeV}$, the minimum Lorentz factor of the GeV emission $\Gamma^{\rm min}_{\rm GeV}$, $E_{iso}$, $E_{\rm LAT}$, and the amount of infalling accreting mass co-rotating (counter-rotating) with the BH $M_{\rm acc}^{\eta_+}$ ($M_{\rm acc}^{\eta_-}$), needed to explaing $E_{\rm LAT}$.}
\label{tab:LAT}
\end{table*}

%%%%%%%%%%%%%%%%%%%%%%%%%%%%%%%%%%%%%%%%%%%%%%%%%%%%%%
\subsection{Reanalyzing the GeV emission of the S-GRB 090227B}\label{sec:5.2}
%%%%%%%%%%%%%%%%%%%%%%%%%%%%%%%%%%%%%%%%%%%%%%%%%%%%%%

We performed the unbinned likelihood analysis method,\footnote{\url{https://fermi.gsfc.nasa.gov/ssc/data/analysis/scitools/lat\_grb\_analysis.html}} which is preferred when the number of events is expected to be small, for the S-GRB 090227B.
We took spectra within $1$~s, $10$~s, $100$~s, and $1000$~s, after the burst trigger.
The background point like sources and diffuse (galactic and extragalactic) emission within $10^{\rm o}$ from the GRB position are taken from LAT 4-year Point Source Catalog (3FGL).
The test statistic (TS) computed from the above likelihood analysis is {\rm TS}$\lesssim1$ in each time interval (${\rm TS}>25$ corresponds to 5-$\sigma$ of significance), therefore, no significant GeV emission can be associated to this GRB.
A single GeV photon with energy $1.59$~GeV at time $896$~s after the trigger and within $1^{\rm o}$ from the GRB has been found.
Considering the above background models, we computed the probability for this photon to belong to this GRB.
The likelihood analysis gives a probability of this photon to correlate to GRB 090227B of $0.36\%$, while its probability of being a photon from the diffuse background is $>99\%$.

The results of this analysis are in agreement with those reported in \citet{Ackermann2013}.
There, it is also stated that an autonomous repoint request by the \textit{Fermi}-GBM brought the LAT down to $\simeq20^{\rm o}$ after $\sim300$~s and, therefore, the source entered in the optimal LAT FoV.
By using the S-GRB common power-law trend $t^{-1.29\pm0.06}$ (see dashed black line in Fig.~\ref{fig:7}), we computed the expected energy fluxes of the GeV emission of the S-GRB 090227B $f_1$, at the time of $\sim300$~s when the source entered the LAT FoV, and $f_2$, at $896$~s when the diffuse background photon was detected. We assumed a power-law spectrum with a typical value of the photon index of $-2$ and obtained $f_1=(1.09\pm0.16)\times10^{-9}$~erg cm$^{-2}$s$^{-1}$ and $f_2=(2.65\pm0.39)\times10^{-10}$~erg cm$^{-2}$s$^{-1}$.
These computed fluxes are within the \textit{Fermi}-LAT sensitivity of the Pass 8 Release 2 Version 6 Instrument Response Functions, 
\footnote{\url{http://www.slac.stanford.edu/exp/glast/groups/canda/lat_Performance_files/broadband_flux_sensitivity_p8r2_source_v6_all_10yr_zmax100_n03.0_e1.50_ts25.png}} which is approximately $10^{-11}$ erg cm$^{-2}$s$^{-1}$.
Therefore, we can conclude that the GeV emission associated to the S-GRB 090227B ceased before $300$~s, when the source entered the LAT FoV.

%%%%%%%%%%%%%%%%%%%%%%%%%%%%%%%%%%%%%%%%%%%%%%%%%%%%%%
\subsection{Lower limits on the GeV emission Lorentz factors in S-GRBs}\label{sec:5.4}
%%%%%%%%%%%%%%%%%%%%%%%%%%%%%%%%%%%%%%%%%%%%%%%%%%%%%%

Following \citet{LitSari2001}, it is possible to derive a lower limit on the Lorentz factor of the GeV emission $\Gamma^{\rm min}_{\rm GeV}$ by requiring that the outflow must be optically thin to high energy photons, namely to the pair creation process.
Using the maximum GeV photon observed energy $E^{\rm max}_{\rm GeV}$ in Tab.~\ref{tab:LAT}, for each S-GRB various lower limits on the GeV Lorentz factors can be derived from the time resolved spectral analysis.
For each S-GRB we estimate lower limits in each time interval of the GeV luminosity light curves in Fig.~\ref{fig:7}. 
Then, $\Gamma^{\rm min}_{\rm GeV}$ for each S-GRB has been then determined as the largest among the inferred lower limits (see Tab.~\ref{tab:LAT}).
The GeV photons are produced in ultrarelativistic outflows with $\Gamma^{\rm min}_{\rm GeV}\gtrsim300$.

%%%%%%%%%%%%%%%%%%%%%%%%%%%%%%%%%%%%%%%%%%%%%%%%%%%%%%
\subsection{The energy budget of the GeV emission in S-GRBs}\label{sec:5.5}
%%%%%%%%%%%%%%%%%%%%%%%%%%%%%%%%%%%%%%%%%%%%%%%%%%%%%%

\citet{2016ApJ...831..178R} proposed that the $0.1$--$100$ GeV in S-GRBs (see Fig.~\ref{fig:7}) is produced by the mass accretion onto the newborn KNBH. 
The amount of mass that remains bound to the BH is given by the conservation of energy and angular momentum from the merger moment to the BH birth.
We can estimate lower limits of the needed mass to explain the energy requirements $E_{\rm LAT}$ in Tab.~\ref{tab:LAT} by considering the above accretion process onto a maximally rotating Kerr BH.
In this case, depending whether the infalling material is in co- or counter-rotating orbit with the spinning BH, the maximum efficiency of the conversion of gravitational energy into radiation is $\eta_+=42.3\%$ or $\eta_-=3.8\%$, respectively (see Ruffini \& Wheeler 1969, in problem $2$ of $\S$~104 in \citealt{LL2003}).
Therefore, $E_{\rm LAT}$ can be expressed as 
\begin{equation}
\label{accretion}
E_{\rm LAT}= f_{\rm b}^{-1}\eta_\pm M_{\rm acc}^{\eta_\pm} c^2\ ,
\end{equation}
where $f_{\rm b}$ is the beaming factor which depends on the geometry of the GeV emission, and $M_{\rm acc}^{\eta_\pm}$ is the amount of accreted mass corresponding to the choice of the efficiency $\eta_\pm$.
The observational evidence that the totality of S-GRBs exhibit GeV emission and that its absence is due instrumental absence of alignment between the LAT and the source
at the time of the GRB emission (see Sec.~\ref{sec:5.2}) suggest that no beaming is necessary in Eq.~\ref{accretion}. 
Therefore, in the following we set $f_{\rm b}\equiv1$.
The corresponding estimates of $M_{\rm acc}^{\eta_\pm}$ in our sample of S-GRBs are listed in Tab.~\ref{tab:LAT}.

%%%%%%%%%%%%%%%%%%%%%%%%%%%%%%%%%%%%%%%%%%%%%%%%%%%%%%
\section{On the detectability of the X-ray emission of S-GRBs }\label{sec:6}
%%%%%%%%%%%%%%%%%%%%%%%%%%%%%%%%%%%%%%%%%%%%%%%%%%%%%%

\begin{figure}
\centering
\includegraphics[width=\hsize,clip]{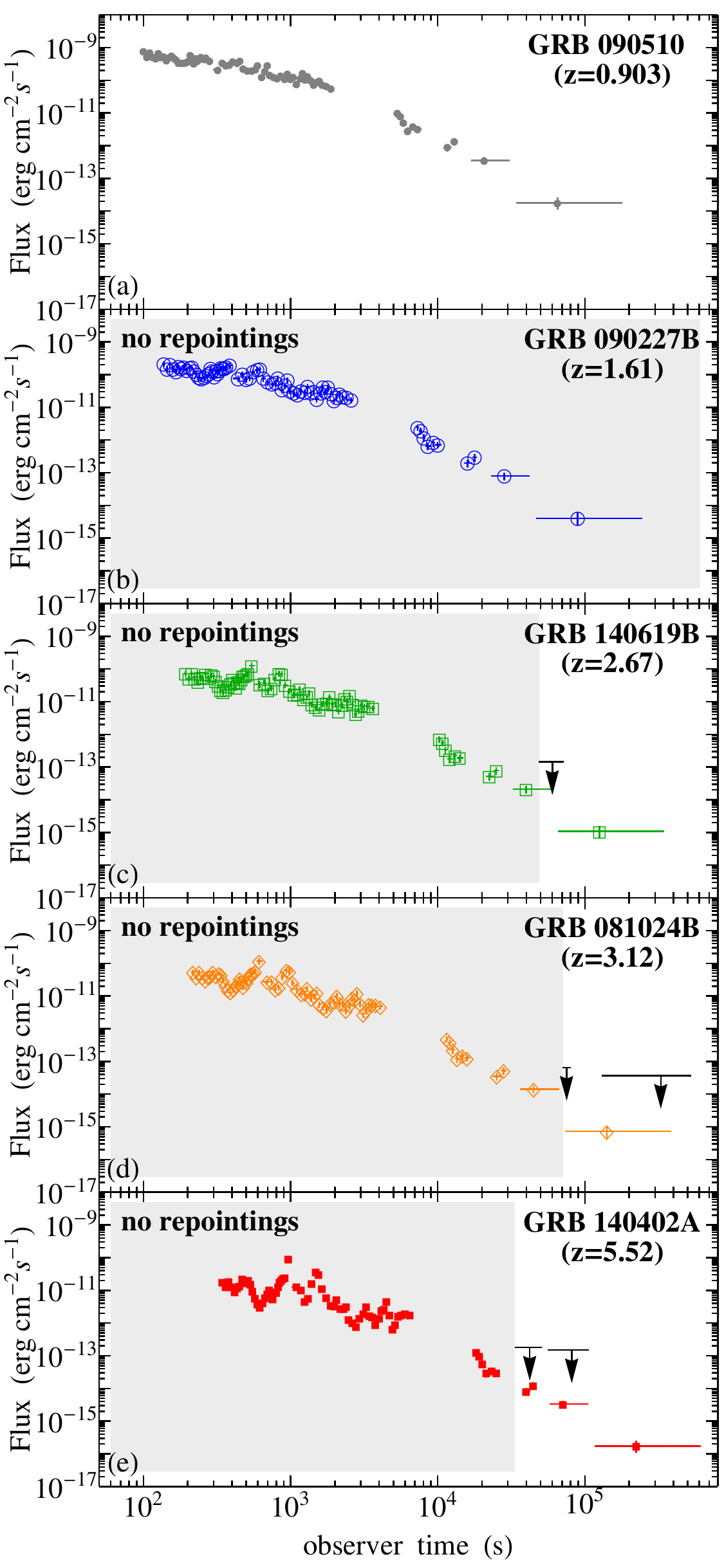}
\caption{The observed $0.3$--$10$ keV energy flux light curves of (a) the S-GRB 090510, located at $z_{\rm in}=0.903$, and the corresponding predicted ones for the S-GRBs (b) 090227B at $z_{\rm in}=1.61$, (c) 140619B at $z_{\rm in}=2.67$, (d) 081024B at $z_{\rm in}=3.12$, and (e) 140402A at $z_{\rm in}=5.52$ (same symbols as in Fig.~\ref{fig:7}. The shaded areas correspond to the epochs before the observational upper limits set by the available \textit{Swift}-XRT repointings (black arrows, see text for details).}
\label{fig:8}
\end{figure}

GRB 090510 is the only S-GRB with a complete X-ray afterglow \citep[see Fig.~\ref{fig:8}(a) and][]{2016ApJ...831..178R}. Only upper limits exist for the X-ray afterglow emission of the other S-GRBs and no special features are identifiable.

As an example to evidence the difficulty of measuring the X-ray afterglow in S-GRBs, we computed the observed X-ray flux light curve of GRB 090510, actually observed at $z_{\rm in}=0.903$, as if it occurred at the redshifts of the other S-GRBs, i.e., $z_{\rm fin}=1.61$, $2.67$, $3.12$, and $5.52$. This can be attained through four steps.
\begin{itemize}
\item[(1)] In each time interval of the X-ray flux light curve $f_{\rm obs}^{\rm in}$ of GRB 090510, we assume that the best fit to the spectral energy distribution is a power-law function with photon index $\gamma$, i.e., $N(E)\sim E^{-\gamma}$. 
\item[(2)] In the rest-frame of GRB 090510, we identify the spectral energy range for a source at redshift $z_{\rm fin}$ which corresponds to the $0.3$--$10$~keV observed by \textit{Swift}-XRT, i.e., 
\begin{equation}
\nonumber 0.3\left(\frac{1+z_{\rm fin}}{1+z_{\rm in}}\right){\rm-}10\left(\frac{1+z_{\rm fin}}{1+z_{\rm in}}\right)\,{\rm keV}\ .
\end{equation}
\item[(3)] We rescale the fluxes for the different luminosity distance $d_{\rm l}$. Therefore, the observed $0.3$--$10$~keV X-ray flux light curve $f_{\rm obs}^{\rm fin}$ for a source at redshift $z_{\rm fin}$ is given by
\begin{align}
\nonumber f_{\rm obs}^{\rm fin}=&f_{\rm obs}^{\rm in}\left[\frac{d_{\rm l}(z_{\rm in})}{d_{\rm l}(z_{\rm fin})}\right]^2\frac{\int^{10\frac{1+z_{\rm fin}}{1+z_{\rm in}}\,{\rm keV}}_{0.3\frac{1+z_{\rm fin}}{1+z_{\rm in}}\,{\rm keV}} N(E)E dE}{\int^{10\,{\rm keV}}_{0.3\,{\rm keV}} N(E) E dE}=\\
=&f_{\rm obs}^{\rm in}\left[\frac{d_{\rm l}(z_{\rm in})}{d_{\rm l}(z_{\rm fin})}\right]^2\left(\frac{1+z_{\rm fin}}{1+z_{\rm in}}\right)^{2-\gamma}.
\end{align}
\item[(4)] We transform the observational time $t_{\rm in}$ of GRB 090510 at $z_{\rm in}$ into the observational time $t_{\rm fin}$ for a source at $z_{\rm fin}$ by taking into account the time dilation due to the cosmological redshift effect, i.e.,
\begin{equation}
t_{\rm fin}=\left(\frac{1+z_{\rm fin}}{1+z_{\rm in}}\right)t_{\rm in}.
\end{equation}
\end{itemize} 

Fig.~\ref{fig:8} shows that all the computed flux light curves are well below the observational upper limits provided by the \textit{Swift}-XRT repointings.
\begin{itemize}
\item[-] S-GRB 090227B, no repointings (see Fig.~\ref{fig:8}(b)).
\item[-] S-GRB 140619B, a repointing from $48.7$ to $71.6$~ks after the GBM trigger with an upper limit of $2.9\times10^{3}$~count/s (see \citealt{GCN16424} and Fig.~\ref{fig:8}(c)).
\item[-] S-GRB 081024B, two repointings within the flux light curve in Fig.~\ref{fig:8}(d). Each upper limit was set by using the lowest count rate among those of the uncatalogued sources within the LAT FoV, later on confirmed as not being the burst X-ray counterparts: the first one at $\sim70.3$~ks after the trigger for $\sim9.9$~ks with a count rate of  $1.3\times10^{-3}$~counts/s \citep{2008GCN..8410....1G}; the second one from $1.5$ to $6.1$~days with an average count rate of $7.4\times10^{-4}$~counts/s \citep{2008GCN..8454....1G}.
\item[-] S-GRB 140402A, two repointings \citep{2014GCN..16078...1P}: the first from $33.3$ to $51.2$~ks with a count rate upper limit of $3.6\times10^{-3}$~counts/s; the second from $56$ to $107$~ks with an upper limit of $3.0\times10^{-3}$~counts/s (see Fig.~\ref{fig:8}(d)).
\end{itemize}
We converted the above count rate upper limits in fluxes by multiplying for a typical conversion factor $5\times10^{-11}$~erg/cm$^2$/counts \citep[see, e.g.,][]{2014GCN..16075...1P}.

We conclude that there is no evidence in favor or against a common behavior of the X-ray afterglows of the S-GRBs in view of the limited observations.

These aspects are noteworthy since in the case of long GRBs the X-ray emission has a very crucial role \citep{2016ApJ...833..159P,2017arXiv170403821R}, which is not testable in the case of S-GRBs.

%%%%%%%%%%%%%%%%%%%%%%%%%%%%%%%%%%%%%%%%%%%%%%%%%%%%%%
\section{On the short bursts originating in BH--NS mergers}\label{sec:usgrb}
%%%%%%%%%%%%%%%%%%%%%%%%%%%%%%%%%%%%%%%%%%%%%%%%%%%%%%

As pointed out in \citet{2015PhRvL.115w1102F}, \citet{2016ApJ...832..136R} and \citet{2016arXiv160203545R}, U-GRBs are expected to originate in the BH--NS binaries produced by the further evolution of the BdHNe \citep[see, e.g.,][]{2016ApJ...833..107B,2016ApJ...832..136R}. We recall that BdHN progenitor systems are composed of a carbon-oxygen core (CO$_{\rm core}$) and a NS in a close binary system. When the CO$_{\rm core}$ explodes as a supernova (SN) Ib/c, its ejecta starts a hypercritical accretion process onto the companion NS, pushing its mass beyond the value $M_{\rm crit}^{\rm NS}$, and leading to the formation of a BH. This BH, together with the new NS ($\nu$NS) produced out of the SN event, leads to the progenitor systems of the U-GRBs.

The orbital velocities of the BH--NS binaries formed from BdHNe are high and even large kicks are unlike to make these systems unbound \citep{2015PhRvL.115w1102F}. U-GRBs represent a new family of BH--NS binaries unaccounted for in current
standard population synthesis analyses \citep[see, e.g.,][]{2015PhRvL.115w1102F}.

U-GRBs are expected to lead to harder and shorter bursts in $\gamma$-rays, which explains the lack of their observational identification \citep{2015PhRvL.115w1102F}, and pose a great challenge possibly to be considered to emit fast radio bursts. They also could manifest themselves, before the merging, as pulsar-BH binaries \citep[see, e.g.,][and references therein]{2015aska.confE..39T}.

%%%%%%%%%%%%%%%%%%%%%%%%%%%%%%%%%%%%%%%%%%%%%%%%%%%%%%
\section{Conclusions}\label{sec:conclusions}
%%%%%%%%%%%%%%%%%%%%%%%%%%%%%%%%%%%%%%%%%%%%%%%%%%%%%%

We have first recalled the division of short bursts into two different sub-classes \citep{2015ApJ...808..190R}: the S-GRFs, with $E_{\rm iso}\lesssim 10^{52}$~erg, $E_{p,i}\lesssim2$~MeV and no GeV emission, and the authentic S-GRBs, with $E_{\rm iso}\gtrsim 10^{52}$~erg, $E_{p,i}\gtrsim2$~MeV and with the presence of the GeV emission, always detected by \textit{Fermi}-LAT, when operative \citep{2015ApJ...808..190R}. 

We then focus on two additional examples of S-GRBs: GRB 081024B, with $E_{iso}=(2.6\pm1.0)\times10^{52}$~erg and $E_{p,i}=(9.6\pm4.9)$~MeV (see Sec.~\ref{sec:2}), and GRB 140402A, with $E_{iso}=(4.7\pm1.1)\times10^{52}$~erg and $E_{p,i}=(6.1\pm1.6)$~MeV (see Sec.~\ref{sec:4}).

We perform time-integrated and time-resolved spectral analyses on both these sources (see Secs.~\ref{sec:2.1.1}--\ref{sec:2.1.2} and Secs.~\ref{sec:4.1.1}--\ref{sec:4.1.2}) and infer their cosmological redshifts ($z=3.12$ for the S-GRB 081024B and $z=5.52$ for the S-GRB 140402A, see Secs.~\ref{sec:2.2.1} and \ref{sec:4.2.1}, respectively).
We also identify their P-GRB spectral emission. The P-GRB emission of S-GRB 081024B exhibit the convolution of BB spectra at different Doppler factors arising from a spinning BH, in total analogy with S-GRB 090510 \citep[see Sec.~\ref{sec:3} and \ref{sec:2.1.2} and][]{2016ApJ...831..178R}. The P-GRB emission of S-GRB 140402A is consistent with a single BB, expected to occur for a moderately spinning BH \citep[see Sec.~\ref{sec:4.1.2}][]{2016ApJ...831..178R}.

The baryon load mass M$_{\rm B}$, the Lorentz $\Gamma$ factor and the properties of the CBM clouds are in agreement with those of the other S-GRBs: M$_{\rm B}\approx10^{-6}$~M$_\odot$, $\Gamma\approx10^4$ (see Secs~\ref{sec:2.2.1} and \ref{sec:4.2.1}), distances of the CBM clouds $r\approx10^{16}$~cm and CBM densities $n_{\rm CBM}\approx10^{-3}$~cm$^{-3}$ (see Secs~\ref{sec:2.2.2} and \ref{sec:4.2.2}), typical of galactic halos environment \citep[see, e.g.,][]{Muccino2012,2015ApJ...808..190R}.

In analogy to the other S-GRBs we confirm that the turn-on of the GeV emission starts after the P-GRB emission and is coeval with the occurrence of the prompt emission (see Sec.~\ref{sec:5}). All these coincidences point to the fact that the GeV emission originates from the on-set of the BH formation \citep[see the space-time diagrams in Fig.~3 of][]{2016ApJ...831..178R}.

Most noteworthy, the existence of a common power-law behavior in the rest-frame $0.1$--$100$~GeV luminosities (see Fig.~\ref{fig:7} in Sec.~\ref{sec:5}), following the BH formation, points to a commonality in the mass and spin of the newly-formed BH in all these S-GRBs. This result is explainable with the expected mass of the merging NSs, each one of $M\approx 1.3$--$1.5 M_\odot$ \citep{2016ARA&A..54..401O}, and the expected range of the non-rotating NS critical mass M$^{\rm NS}_{\rm crit}\sim 2.2$--$2.7$~M$_\odot$ leading to a standard value of the BH mass and of its Kerr parameter $a/M\sim1$ \citep{2015ApJ...808..190R}.

Finally, in all S-GRBs the energetic of the GeV emission implies the accretion of $M\gtrsim0.03$--$0.08 M_\odot$ or $M\gtrsim0.35$--$0.86 M_\odot$ for co- or counter-rotating orbits with a maximally rotating BH, respectively (see Sec.~\ref{sec:5}). This accretion process, occurring both in S-GRBs and also BdHNe \citep{2016ApJ...833..107B}, is currently being analyzed for the occurrence of r-process \citep{Ruffini2014,2016ApJ...833..107B}.

In all the identified S-GRBs, within the Fermi-LAT FoV, GeV photons are always observed \citep{2016ApJ...831..178R,2016ApJ...832..136R}. 
This implies that no intrinsic beaming is necessary for the S-GRB GeV emission.
The Lorentz factor of the GeV emission is $\Gamma^{\rm min}_{\rm GeV}\gtrsim300$.

\begin{figure}
\centering
\includegraphics[width=\hsize,clip]{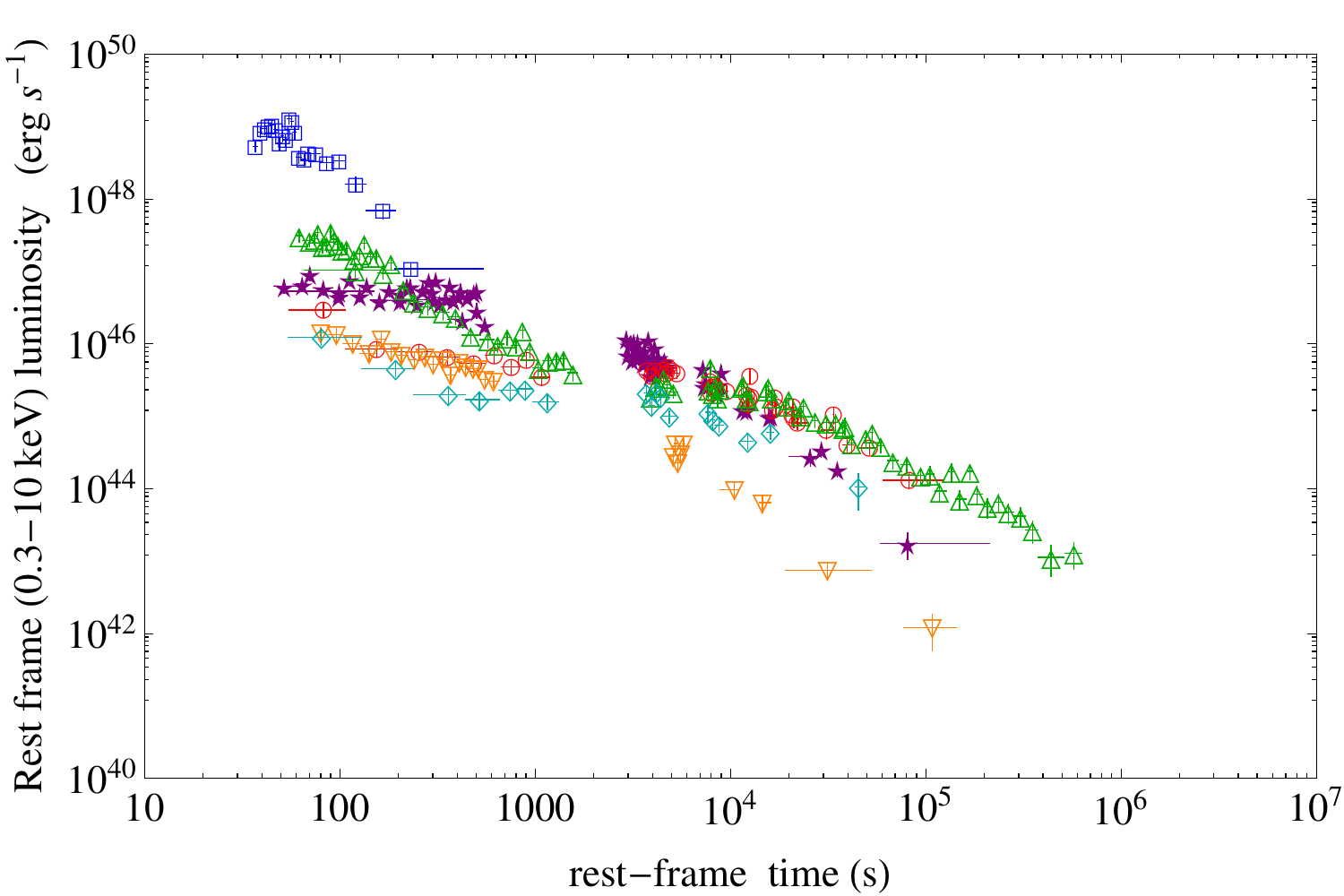}
\caption{The rest-frame $0.3$--$10$~keV X-ray luminosity light curves of selected S-GRFs: GRB 051210 (blue squares), GRB 051221 (green triangles), GRB 061201 (orange reversed triangles), GRB 070809 (cyan diamonds), and GRB 130603B (purple stars).}
\label{fig:SGRFs}
\end{figure}

From Fig.~\ref{fig:8} for the S-GRBs and from Fig.~\ref{fig:SGRFs} for S-GRFs we conclude that in both systems there is no evidence for the early X-ray flares observed in BdHNe \cite{2017arXiv170403821R}.

Before closing, we return to the issue of the GW detectability by aLIGO from S-GRBs. We have already evidenced their non detectability in GRB 090227B \citep{Oliveira2014} and GRB 140619B \citep{2015ApJ...808..190R} by aLIGO by computing the signal to noise ratio S/N up to the contact point of the binary NS components. In both cases each NS has been assumed to have mass $M_{\rm NS}=1.34$~$M_\odot=0.5M_{\rm crit}^{\rm NS}$. There, it has been concluded that the GW signals emitted in such systems were well below the value S/N$=8$ needed for a positive detection.

These considerations have been extended in \citet{2016arXiv160203545R} to all S-GRBs. It was there concluded that such signals might be detectable for sources located at $z\lesssim 0.14$ (i.e., at distances smaller than the GW detection horizon of $640$~Mpc) for the aLIGO $2022+$ run. GRB 090510, to date the closest S-GRB, is located at $z=0.903$ (i.e., $5842$~Mpc) and, therefore, it is outside such a GW detection horizon. 
We can then conclude that for sources at distances larger than that of GRB 090510, like GRB 081024B (at $z=3.12$) and GRB 140402A (at $z=5.52$) analyzed in this paper, no GW emission can be detected.

\acknowledgments

We thank the referee for pleasant and expert advices. M.~M. and J.~A.~R. acknowledge the partial support of the project N 3101/GF4 IPC-11, and the target program F.0679  0073-6/PTsF of the Ministry of Education and Science of the Republic of Kazakhstan. Y.~A. is  supported by the Erasmus Mundus Joint Doctorate Program by Grant Number  2014-0707  from the agency EACEA of the European Commission.

\end{document}